\documentclass[a4paper,11pt]{article}
\pdfoutput=1 

\usepackage{jcappub}
\usepackage{slashed}
\usepackage{multirow}
\usepackage{color}
\usepackage[normalem]{ulem}
\usepackage[T1]{fontenc} 
\newcommand{\be}{\begin{equation}}
\newcommand{\ee}{\end{equation}}

\newcommand{\dN}{\Delta N_\text{eff}}

\title{\boldmath Hot Axions and the $H_0$ tension}

\author[a,b]{Francesco D'Eramo,}
\author[c]{Ricardo Z. Ferreira,}
\author[c]{Alessio Notari,}
\author[c]{Jos\'e Luis Bernal}

\affiliation[a]{Dipartimento di Fisica ed Astronomia, Universit\`a di Padova, Via Marzolo 8, 35131 Padova, Italy}
\affiliation[b]{INFN, Sezione di Padova, Via Marzolo 8, 35131 Padova, Italy}
\affiliation[c]{Departament de F\'isica Qu\`antica i Astrofis\'ica \& Institut de Ci\`encies del Cosmos (ICCUB), Universitat de Barcelona, Mart\'i i Franqu\`es 1, 08028 Barcelona, Spain}

\emailAdd{francesco.deramo@pd.infn.it}
\emailAdd{rferreira@icc.ub.edu}
\emailAdd{notari@fqa.ub.edu}
\emailAdd{joseluis.bernal@icc.ub.edu}

\abstract{Scattering and decay processes of thermal bath particles involving heavy leptons can dump hot axions in the primordial plasma around the QCD phase transition. We compute their relic density, parameterized by an effective number $\Delta N_{\rm eff}$ of additional neutrinos. For couplings allowed by current bounds, production via scattering yields $\dN \lesssim 0.6$ and $\dN \lesssim 0.2$ for the cases of muon and tau, respectively. Flavor violating tau decays to a lighter lepton plus an axion give $\Delta N_{\rm eff} \lesssim 0.3$. Such values of $\dN$ can alleviate the tension between the direct local measurement of the Hubble constant $H_0$ and the inferred value from observations of the Cosmic Microwave Background, assuming $\Lambda$CDM. We analyze present cosmological data from the Planck collaboration and baryon acoustic oscillations with priors given in terms of the axion-lepton couplings. For axions coupled to muons, the tension can be alleviated below the 3$\sigma$ level. Future experiments will measure $\Delta N_{\rm eff}$ with higher precision, providing an axion discovery channel and probing the role of hot axions in the $H_0$ tension.}

\begin{document}
\maketitle
\flushbottom

\section{Introduction}
\label{sec:intro}

Axion-like particles (ALPs) are motivated candidates for extremely light and weakly-coupled degrees of freedom beyond the Standard Model (SM). The motivation is notably robust for the QCD axion, as the Peccei-Quinn (PQ) mechanism is an elegant solution to the strong CP problem~\cite{Peccei:1977hh,Peccei:1977ur} and it could account for the observed dark matter (DM) abundance~\cite{Wilczek:1977pj,Weinberg:1977ma,Kim:1979if,Shifman:1979if,Zhitnitsky:1980tq,Dine:1981rt,Preskill:1982cy,Abbott:1982af,Dine:1982ah}. Axion phenomenology is quite broad, detection strategies are multiple and complementary. Direct interactions with SM particles, unavoidably present for the QCD axion, are the subject of present and future searches~\cite{Arik:2008mq,Irastorza:2011gs,Graham:2013gfa,Budker:2013hfa,Irastorza:2013dav,Armengaud:2014gea,Kahn:2016aff,Barbieri:2016vwg,Melcon:2018dba,Du:2018uak}. 

This paper focuses on a peculiar cosmological imprint: scatterings and decays of thermal bath particles can produce relativistic axions throughout the expansion history of our universe~\cite{Turner:1986tb,Berezhiani:1992rk,Masso:2002np,Graf:2010tv,Brust:2013xpv,Salvio:2013iaa,Baumann:2016wac,Ferreira:2018vjj}. The resulting growth in the radiation energy density, traditionally parameterized by $\dN$ additional neutrino species, can be probed by observations of the Cosmic Microwave Background (CMB) and Baryon Acoustic Oscillations (BAO)~\cite{doi:10.1080/00107514.2013.818063, Lesgourgues:2014zoa, Archidiacono:2013fha}. The current best-fit value from the latest Planck 2018+BAO data \cite{Planck2018_cosmopars}, $N_\text{eff}=2.99 \pm 0.17$, is in agreement with the SM prediction, $N_\text{eff(SM)} = 3.046$~\cite{Dodelson:1992km,Hannestad:1995rs,Dolgov:1997mb,Mangano:2005cc,deSalas:2016ztq}. Ongoing and forthcoming experimental efforts make the present time remarkably exciting. Current experiments~\cite{SPT-3G} will reach a sensitivity $\dN\sim 0.06$ in the near future, whereas future CMB-S4 surveys forecast an improvement down to $\dN\sim 0.024$~\cite{Abazajian:2016yjj}. As a useful benchmark, a $2 \sigma$ detection with the latter sensitivity would require hot axions that were in thermal equilibrium at early times and then decoupled at temperatures $T \lesssim 10 \, {\rm GeV}$~\cite{Ferreira:2018vjj}. Combined with the complementary information from direct searches, this makes the next few years very promising for the quest for axions. 

Recently, a mismatch between low and high redshift determinations of the Hubble constant $H_0$ triggered further interest in $\dN$. The value inferred from CMB observations (assuming $\Lambda$CDM), $H_0=67.27\pm 0.60$  km $\text{s}^{-1}$ Mpc$^{-1}$~\cite{Planck2018_cosmopars}, is in $3.6\sigma$  tension with direct measurements from supernovae, $H_0= 73.52 \pm 1.62$ km $\text{s}^{-1}$ Mpc$^{-1}$  \cite{Riess:2018byc}. The latter also include data from the latest GAIA release \cite{2018arXiv180409365G}. Such a tension, although weaker, also appears in other independent measurements of $H_0$ that probe the local universe and observations at high redshift (see, {\it e.g.}, Ref.~\cite{Bernal_baccus,Dhawan:2017ywl}). Notably, BAO are a complementary probe that depends on an assumed expansion history of the Universe and on the product $r_{\rm s} h$~\cite{Aubourg_2015,StandardQuantities}, where $r_{\rm s}$ and $h$ are the sound horizon at radiation drag and the reduced Hubble constant $h=H_0/100$ km $\text{s}^{-1}$ Mpc$^{-1}$, respectively. An independent determination of $r_{\rm s}$ would make it possible to infer the value of $H_0$ using BAO. Measurements of $H_0$ from BAO using $r_{\rm s}$ from the CMB \cite{Alam:2016hwk,Zarrouk:2018vwy} or from primordial deuterium measurements~\cite{Cooke_deu2017}, as in \cite{Addison_bao13,Addison_H0bao}, are also in tension with the direct local measurement. The combination of CMB and BAO leads to a 3.46$\sigma$ disagreement. The tension persists even if we parameterize the cosmic expansion at low redshifts in a model independent way~\cite{Bernal:2016gxb}, since supernova type~Ia measurements (up to $z\sim 1$) force $H(z)/H_0$ to be $\Lambda$CDM-like with deviations $ \lesssim 5\%$ at low redshift. Exotic dark energy models that change the expansion history at low redshift to reconcile the two $H_0$ measurements seem disfavored (see also~\cite{Mortsell:2018mfj}). Remarkably, having $\dN>0$ is a way to decrease the tension~\cite{RiessH0_2016,Bernal:2016gxb}; a combined fit~\cite{Planck2018_cosmopars} with the new 2018 Planck data, including lensing, BAO and the direct $H_0$ measurement~\cite{2018ApJ...855..136R} leads to $N_{\rm eff}=3.27\pm0.15$ and $H_0=(69.32\pm0.97) {\rm \, km \, s^{-1} \, Mpc^{-1}}$.

Motivated by the above considerations, we study thermal production of axions coupled to heavy leptons. This mechanism is mostly active at temperatures around the lepton mass~\cite{Turner:1986tb}, hence close to the QCD phase transition (QCDPT).~\footnote{Bounds on axion-electron couplings~\cite{Viaux:2013lha} lead to unobservably small  $\dN$. Couplings to heavy quarks ($c$, $b$ and $t$) give $\dN$ observable in the near future~\cite{Ferreira:2018vjj}, whereas couplings to light quarks ($u$, $d$ and $s$) require a treatment of hadronic bound states.} The resulting comoving relic abundance is quite large, since it is inversely proportional to the number of relativistic degrees of freedom that sharply decreases around this epoch. The consequence is of a twofold nature: future CMB experiments can observe axions; hot axions produced via this mechanism can alleviate the current $H_0$ tension. An estimate of $\dN$ via these couplings was provided in Refs.~\cite{Turner:1986tb,Brust:2013xpv,Baumann:2016wac}. Here, we perform a careful analysis by computing the full cross-sections and solving the Boltzmann equations to compute the precise value of $\dN$. Our results are valid also when a complete thermalization is never reached. With these results in hand, we re-analyze CMB and BAO data to reassess the $H_0$ tension in light of this theoretical framework.

We present axion interactions with heavy leptons in Sec.~\ref{Lag}. After establishing the experimental bounds on such interactions, we compute in Sec.~\ref{Sec:production} the axion relic density for allowed couplings. The effect of such a hot axion population is parameterized by a number $\dN$ of additional neutrinos, as we quantify in Sec.~\ref{sec:dN}. Up to this point, we only assume axion derivative coupling to leptons. In Sec.~\ref{sec:QCD} we discuss the results in the context of the QCD axion. Then, we study how these hot axions and their theory based priors affect the tension in the measured value of $H_0$ in Sec.~\ref{sec:cosmo}, and we give our conclusions in Sec.~\ref{sec:Conclusion}. We also provide detailed appendices with our calculations.

\section{Axion Couplings}
\label{Lag}

We introduce the operators that couple the axion to the visible sector and identify the couplings not excluded by current experiments. Within this allowed region, we compute the relic density and the consequent contribution to $\dN$ in Secs.~\ref{Sec:production} and \ref{sec:dN}, respectively. 

\subsection{Lagrangian}

The Lagrangian for the class of theories considered in this work takes the form
\be
\mathcal{L} = \mathcal{L}_{\rm SM} + \frac{1}{2} (\partial_\mu a)^2 - \frac{1}{2} m_a^2 a^2 + \mathcal{L}^{(a)}_{\rm int} \ .
\label{eq:Lgeneral}
\ee
In addition to the SM part and the axion kinetic and mass terms, we have axion effective interactions with SM fields. The axion field $a$ can be thought as a light degree of freedom arising from the spontaneous breaking of a global symmetry. If the symmetry is broken only by the vacuum, the axion field $a$ is a Goldstone boson and the low energy effective Lagrangian must be invariant under the continuous shift symmetry $a \rightarrow a + c$, with $c$ an arbitrary real constant. We immediately notice how the mass term breaks such a shift symmetry, thus the axion is a pseudo-Goldstone boson for $m_a \neq 0$. 

We parameterize the interactions as follows
\be
\mathcal{L}^{(a)}_{\rm int} =  \frac{1}{2f}  \, \partial_\mu a \, J_a^\mu  + \frac{a}{f} \sum_X  \frac{ \alpha_X}{8 \pi} C_{XX} X_{\mu \nu} \tilde{X}^{\mu \nu} \ .
\label{eq:Lint}
\ee
All operators are suppressed by the dimensionful parameter $f$, which we identify as the scale of spontaneous breaking for the global symmetry giving the axion. Such a parameter is also the cutoff scale for the effective theory considered here. The first type of interactions couples the axion space-time derivative with the spin-1 current $J_a^\mu$ built out of SM matter fields, whereas the second class of operators couples the axion to gauge bosons. Here, the index $X$ runs over the SM gauge group, with $X_{\mu\nu}$ and $\alpha_X = g_X^2 / (4 \pi) $ the associated field strength and gauge coupling constant, respectively. While the former are still invariant under the generic shift $a \rightarrow a + c$, the same is not true for the latter; the total Lagrangian is invariant up to a total space-time derivative. This extra term is irrelevant for axion couplings to Abelian gauge bosons (such as the photon, $X = \gamma$). The same conclusion does not hold for an axion coupling to gluons  ($X = G$). Upon using the conventional normalization $C_{GG} = 1$, the coupling with gluons is only invariant under the discrete shift $a \rightarrow a + 2 \pi f$.

This paper focuses on axion production via processes involving leptons.  We decompose the leptonic currents into diagonal and off-diagonal parts in flavor space
\be
J_a^\mu = \left.J_a^\mu\right|_{\rm diag} + \left.J_a^\mu\right|_{\rm off-diag}  \ , 
\label{eq:axioncurrent}
\ee
Vector currents are conserved for diagonal couplings, hence only axial-vector currents give a physical effect~\footnote{True also for virtual effects if we limit to Green functions up to $\mathcal{O}(1/f)$ in the power counting~\cite{Arzt:1993gz}.}
\be
\left.J_a^\mu\right|_{\rm diag} = \sum_{\ell = e, \mu, \tau} c_\ell \, \bar{\ell} \gamma^\mu \gamma^5 \ell \ .
\label{eq:currentdiagonal}
\ee
On the contrary, we consider both vector and axial-vector currents for the off-diagonal part
\be
\left.J_a^\mu\right|_{\rm off-diag} = \sum_{\ell \neq \ell^\prime} \bar{\ell^\prime} \gamma^\mu \left(\mathcal{V}_{\ell^\prime \ell} + \mathcal{A}_{\ell^\prime \ell}  \gamma^5 \right) \ell  + {\rm h.c.} \ ,
\ee
 where we can assume the matrix coefficients $\mathcal{V}_{\ell^\prime \ell}$ and $\mathcal{A}_{\ell^\prime \ell}$ to be real, since we will always use them for tree level calculations.
 
The effective theory in Eq~\eqref{eq:Lgeneral} with the interactions given in Eq~\eqref{eq:Lint} has several free parameters. We identify three benchmark theories we are interested in, defined as follows.
\begin{itemize}
\item \textbf{Leptophilic ALP:} The axion interacts only with leptons. This could correspond to the case where the  global symmetry whose breaking yields $a$ is broken only by the vacuum, in which case the axion is a true Goldstone boson. Even a small explicit breaking due to $m_a \neq 0$ would fit within this class of theories, as long as interactions with gauge bosons are still forbidden at the cutoff scale. As discussed below, radiative corrections generate these interactions~\cite{Bauer:2017ris} with a scaling $C_{\gamma \gamma}\approx c_\ell \, m^2_a / m^2_\ell$. Providing a small amount of the shift symmetry breaking, the axion mass is much smaller than the lepton masses and so the loop-induced coupling can be neglected.
\item \textbf{Lepto and Photophilic ALP:} Along with lepton couplings, we also allow order one interactions with photons. Note how, even if we assume $C_{\gamma\gamma} \simeq \mathcal{O}(1)$, the interaction with photons is suppressed by the loop factor $\alpha_{em} / (8 \pi)$ with respect to the one with leptons. Although this does not affect our cosmological study, the additional coupling with photons are probed by axion searches~\cite{Arik:2008mq,Irastorza:2011gs,Irastorza:2013dav,Armengaud:2014gea, Melcon:2018dba} and they can potentially exclude regions of the parameter space.
\item \textbf{QCD Axion:} The axion field arises from the breaking of a PQ symmetry, with consequent interactions with gluons. We employ the conventional normalization $C_{GG} = 1$ in Eq.~\eqref{eq:Lint}. Barring finely-tuned cancellations between UV and IR contributions, interactions with photons $C_{\gamma\gamma} \simeq \mathcal{O}(1)$ are also generically present. Moreover, the coupling with gluons is a source for an axion mass term due to QCD instantons.
\end{itemize}
The first class of theories has the minimal ingredients needed for our cosmological study. As shown in Sec.~\ref{Sec:production},  axion production in the early universe via lepton scattering and decay is mostly active at temperatures around the mass of the heaviest lepton involved in the process. Given the numerical values for our cases ($m_\tau \simeq 1.77 \, {\rm GeV}$ and $m_\mu \simeq 105 \, {\rm MeV}$), axions are produced via these channels at quite low temperatures around or below the QCDPT. On the contrary, production via gauge boson scattering is peaked at high temperatures~\cite{Masso:2002np,Graf:2010tv,Salvio:2013iaa} and it is a subdominant contribution at the QCDPT. For this reason, we focus on axion production via processes involving leptons in Secs.~\ref{Sec:production} and \ref{sec:dN}. The impact of having an order one coupling to photons or gluons is deferred to Sec.~\ref{sec:QCD}, where we discuss complementary constraints within the context of the QCD axion.

\subsection{Experimental Constraints on Axion-Lepton Couplings \label{Sec:constr}}

Before exploring early universe physics, we identify the phenomenologically viable couplings. A summary of these constraints can be found in table 1 of Ref.~\cite{Baumann:2016wac}. Here, we add a discussion on the applicability of some of such bounds, such as the model dependence of the loop induced constraints and the robustness of the supernova bounds on the axion-muon coupling.

A direct coupling to electrons is strongly constrained by stellar cooling~\cite{Raffelt:1994ry,Viaux:2013lha,Patrignani:2016xqp}
\be \label{c_e}
f / c_e \gtrsim 5 \times 10^9 \, {\rm GeV} \ .
\ee
This case is uninteresting for our purposes, since $\dN$ for such a large value of $f / c_e$ is negligible. For this reason, we assume this coupling to be much smaller than the ones to heavier leptons.

Stellar physics can also bound couplings to muons. However, muons are too heavy to be produced in main sequence stars and one has to consider hotter environments in order to derive meaningful bounds. Ref.~\cite{Brust:2013xpv} found $f/c_\mu \gtrsim 6 \times 10^6 \, {\rm GeV} $ by applying the constraint from the supernova (SN) 1987A explosion with $T\simeq 30$ MeV. We revisit this constraint in App.~\ref{App:Axion-muon}, showing how the numerical bound has a strong dependence on the supernova temperature; as we vary the SN temperature in the window $20-60$ MeV, the lower limit on $f/c_\mu$ spans the range $10^6-10^9$ GeV. Another legitimate concern is the assumption of a muon thermal spectrum during the explosion; the supernovae temperature is always much smaller than the muon mass, which in turn implies a large uncertainty in the initial muon abundance.

A more robust bound arises from radiative corrections. Indeed, interactions with electrons are induced at one-loop if the axion couples with a heavier lepton at tree-level. We defer the details of the calculation to App.~\ref{app:loop} and we just quote the results here 
\begin{eqnarray}
\label{eq:loopboundtau} f/c_\tau & \gtrsim &7 \times 10^4 \, {\rm GeV} \, , \\
\label{eq:loopboundmu} f/c_\mu & \gtrsim & 200  \, {\rm GeV} \ .  
\end{eqnarray}

The correction is proportional to $m^2_\ell$ and it is more relevant for the tau-philic case. This constraint is model dependent as it relies on the evolution  of the axion-electron coupling with the energy scale, which depends on the UV details of the model. Finally, a derivative coupling to leptons also induces a $C_{\gamma \gamma}\approx c_\ell \, m^2_a / m^2_\ell$~\cite{Bauer:2017ris}, and this is always negligible ($C_{\gamma \gamma} \ll 10^{-16}$) for $m_a\ll {\cal O} (\text{eV})$.

We conclude with bounds on the non-diagonal couplings from rare lepton decays~\cite{Bolton:1988af,Albrecht:1995ht}
\begin{eqnarray}
f/c_{\mu e}&\gtrsim&3\times 10^9 \,\text{GeV} \, ,\\
f/c_{\tau e} &\gtrsim&4 \times 10^6 \,\text{GeV} \, , \\ 
f/c_{\tau \mu}&\gtrsim&3\times 10^6 \, \text{GeV}\,.
\end{eqnarray}
Here, we define the effective coupling $c_{\ell \ell'}\equiv \sqrt{\mathcal{V}^2_{\ell^\prime \ell} + \mathcal{A}^2_{\ell^\prime \ell}}$. We will focus on $c_{\tau e}$ and  $c_{\tau \mu}$, as the above constraints make the $c_{\mu e}$ case uninteresting.

\section{Production of Hot Axions in the Early Universe} 
\label{Sec:production}

Scattering and decay processes of thermal bath particles dump hot axions in the early universe. The number density $n_a$ of axions can be tracked down by solving the Boltzmann equation
\be
\frac{d n_a}{d t} + 3 H n_a = \left(\sum_S \overline{\Gamma}_{S} + \sum_D \overline{\Gamma}_{D}\right) \left(n_a^{\rm eq} - n_a \right) \ .
\label{Boltz}
\ee
Here, $n_a^{\rm eq}$ is the axion equilibrium number density. The dilution due to the expansion is captured on the left-hand side by the Hubble parameter
\be
H = \frac{1}{\sqrt{3} M_{\rm Pl}} \sqrt{\rho} =  \frac{1}{\sqrt{3} M_{\rm Pl}} \left( \frac{\pi^2}{30} g_* T^4 \right)^{1/2} \ ,
\label{eq:Hubble}
\ee
with the last expression valid for a radiation dominated universe with $g_*$ relativistic degrees of freedom. The effect of number changing processes is captured by the right-hand side. 

For a generic scattering process $B_1 B_2 \, \rightarrow \, B_3 a$, with $B_i$ thermal bath degrees of freedom, the associated rate reads~\cite{Gondolo:1990dk,Giudice:2003jh,DEramo:2017ecx}
\be
\overline{\Gamma}_{S} = \frac{n_1^{\rm eq} n_2^{\rm eq}}{n_a^{\rm eq}} \langle \sigma_{B_1 B_2 \, \rightarrow \, B_3 a} v_{\rm rel} \rangle
 \ . 
\label{eq:collscattering}
\ee
The cross section $\sigma_{B_1 B_2 \, \rightarrow \, B_3 a}$ is multiplied by the Moeller velocity $v_{\rm rel}$, and such a product is averaged over a thermal ensemble. For production via bath particles decays $B_1 \rightarrow B_3 a$, the associated expression for the rate reads~\cite{Hall:2009bx,DEramo:2017ecx}
\be
\overline{\Gamma}_{D}  = \frac{n_1^{\rm eq}}{n_a^{\rm eq}} \, \Gamma_{B_1 \, \rightarrow \, B_3 a}    \, \frac{K_1\left[ m_1 / T\right]}{K_2\left[m_1 / T\right]} \ , 
\label{eq:colldecay}
\ee
where $\Gamma_{B_1 \, \rightarrow \, B_3 a}$ is the rest frame decay width and $K_1$ and $K_2$ are modified Bessel function of the second kind. The cross section appearing in Eq.~\eqref{eq:collscattering} and the decay width in Eq.~\eqref{eq:colldecay} are evaluated by averaging and summing over initial and final states, respectively. Multiplicity factors (accounting for helicity, colors, etc.) appear in the number density functions. Explicit expressions for cross sections, decay widths, equilibrium densities $n_i^{\rm eq}$ and details about the thermal average procedure can be found in App.~\ref{app:calculations}.  

We introduce the dimensionless axion comoving number density $Y_a = n_a / s$, in so doing scaling out the effect of the Hubble expansion. The entropy density reads $s\equiv 2 \pi^2 g_{*s} T^3/45$, with $g_{*s}$ not the same as $g_*$ (the one in Eq.~\eqref{eq:Hubble})\footnote{The number of relativistic degrees of freedom $g_*$ and the number of entropic degrees of freedom $g_{*s}$ evolve with the temperature, respectively, as \begin{eqnarray}
	g_*= \sum_\text{i} \beta_i \,g_i  \left(\frac{T_i}{T}\right)^4 \qquad , \qquad
	g_{*s}= \sum_\text{i} \beta_i \, g_i \left(\frac{T_i}{T}\right)^3
	\end{eqnarray} where $\beta_i=1$ for bosons and $7/8$ for fermions, $T$ is the photon temperature, $T_i$ the temperature of each species and the summation runs over thermalized and decoupled species.}. A convenient ``time variable'' is the inverse temperature $x = m_\ell / T$, with $\ell$ the heaviest lepton involved in the processes under consideration. In terms of these variables, the Boltzmann equation reads
\be
s H x \frac{d Y_a}{dx}  =  \left( 1-\frac{1}{3} \frac{d \ln g_{*s}}{d\ln x} \right) \left(\sum_S \gamma_{S} + \sum_D \gamma_{D}\right)  \left(1-\frac{Y_a}{Y_a^{\rm eq}}\right) \ ,
\label{eq:BEmain}
\ee
where $Y^{\rm eq}_a =  n^{\rm eq}_a / s$ and we define the effective rates $\gamma_{S, D} \equiv n_a^{\rm eq} \, \overline{\Gamma}_{S, D}$.

It is instructive to derive an approximate analytical solution, valid upon neglecting the temperature dependence of $g_*$ and $g_{*s}$~\cite{Ferreira:2018vjj}. Within this approximation, the Hubble parameter and the entropy density scale as $H = H(m_\ell) \,x^{-2}$ and $s \equiv s(m_\ell) \,x^{-3}$, whereas the rates for scattering and decay approximately scale as $\gamma_S \equiv \gamma_S(m_\ell) \,x^{-4} e^{-x}$ and $\gamma_D \equiv \gamma_D(m_\ell) \,x^{-3} e^{-x}$, respectively. Once we impose $Y_a(x=0)=0$ as the initial condition, we find the solution
\begin{eqnarray}
Y_a(x) \simeq Y_a^{\rm eq} \left[1-e^{-(1-e^{-x}) (r_D+r_S)+x e^{-x}r_D }\right] \ ,
\label{eq:Yanalytical}
\end{eqnarray} 
where $ r_{S,D} \equiv ( \overline{\Gamma}_{S,D}/H)_{T=m_\ell}$. The asymptotic value at small temperatures, which as explained in the next section determines the effective number of additional neutrinos, results in
\be
Y_a(x=\infty) \simeq Y_a^{\rm eq} \left[1-e^{-(r_S+r_D)}\right]  \ . 
\label{semianalytical}
\ee

Before presenting numerical results, it is worth expanding our discussion on the temperature dependence of $g_*$ and $g_{*s}$. Other than the logarithmic derivative appearing in Eq.~\eqref{eq:BEmain}, the shape of $g_*$ across the QCD phase transition also affects the temperature dependence of the function $Y_a^{\rm eq}$. Given the value of the tau and muon masses, processes producing axions are mostly active around the QCD phase transition, so the temperature dependence of $g_*$ and $g_{*s}$ can potentially have an impact on our results. We numerically solve the Boltzmann equation by using the two different results for $g_*$ and $g_{*s}$ from Refs.~\cite{Laine:2006cp} and \cite{Borsanyi:2016ksw}. The former provides a calculation within the framework of dimensionally reduced effective field theories, whereas the latter relies on lattice simulations. We take the range between the results obtained by employing the two different $g_*$ and $g_{*s}$ as the theoretical uncertainty of our prediction. 

\subsection{Axion Production for Diagonal Interactions}

Kinematics only allow axion production via scattering for flavor diagonal interactions. Lepton pair annihilation can give an axion and a photon in the final state. Alternatively, leptons (or anti-leptons) can collide with a photon and produce an axion. The two processes read
\be
\ell^+ \ell^ - \, \rightarrow \, \gamma a \ .
\ee
\be
\ell^\pm \gamma \, \rightarrow \, \ell^\pm a \ .
\ee
The total rate, accounting for both contributions, results in
\be
\sum_S \overline{\Gamma}_{S} =  \frac{n_{\ell^+}^{\rm eq} n_{\ell^-}^{\rm eq}}{n_a^{\rm eq}} \langle \sigma_{\ell^+ \ell^ - \, \rightarrow \, \gamma a} v_{\rm rel} \rangle + 2 \times \frac{n_{\ell^-}^{\rm eq} n_{\gamma}^{\rm eq}}{n_a^{\rm eq}} \langle \sigma_{\ell^- \gamma \, \rightarrow \, \ell^- a} v_{\rm rel} \rangle \ .
\ee
The factor of $2$ in front of the second contribution accounts for production via both lepton and anti-lepton (this is justified since at $T\approx m_\ell$ we approximately have $n_{\ell^-}^{\rm eq} = n_{\ell^+}^{\rm eq}$) . 

\subsection{Axion Production for Off-Diagonal Interactions}

With off-diagonal leptonic currents, direct decays are kinematically allowed
\be
\ell^\pm \, \rightarrow \, \ell^{\prime \pm}  a \ .
\ee
The rate for these processes reads
\be
\overline{\Gamma}_{D}  = 2 \times \frac{n_{l^-}^{\rm eq}}{n_a^{\rm eq}} \, \Gamma_{\ell^- \, \rightarrow \, \ell^{\prime -}  a}    \, \frac{K_1\left[ m_\ell / T\right]}{K_2\left[m_\ell / T\right]} \ ,
\ee
where the factor of $2$ accounts again for decays of both leptons and anti-leptons. 

\begin{figure}
	\begin{center}
		\vspace*{3mm}
		\includegraphics[width=0.8\linewidth]{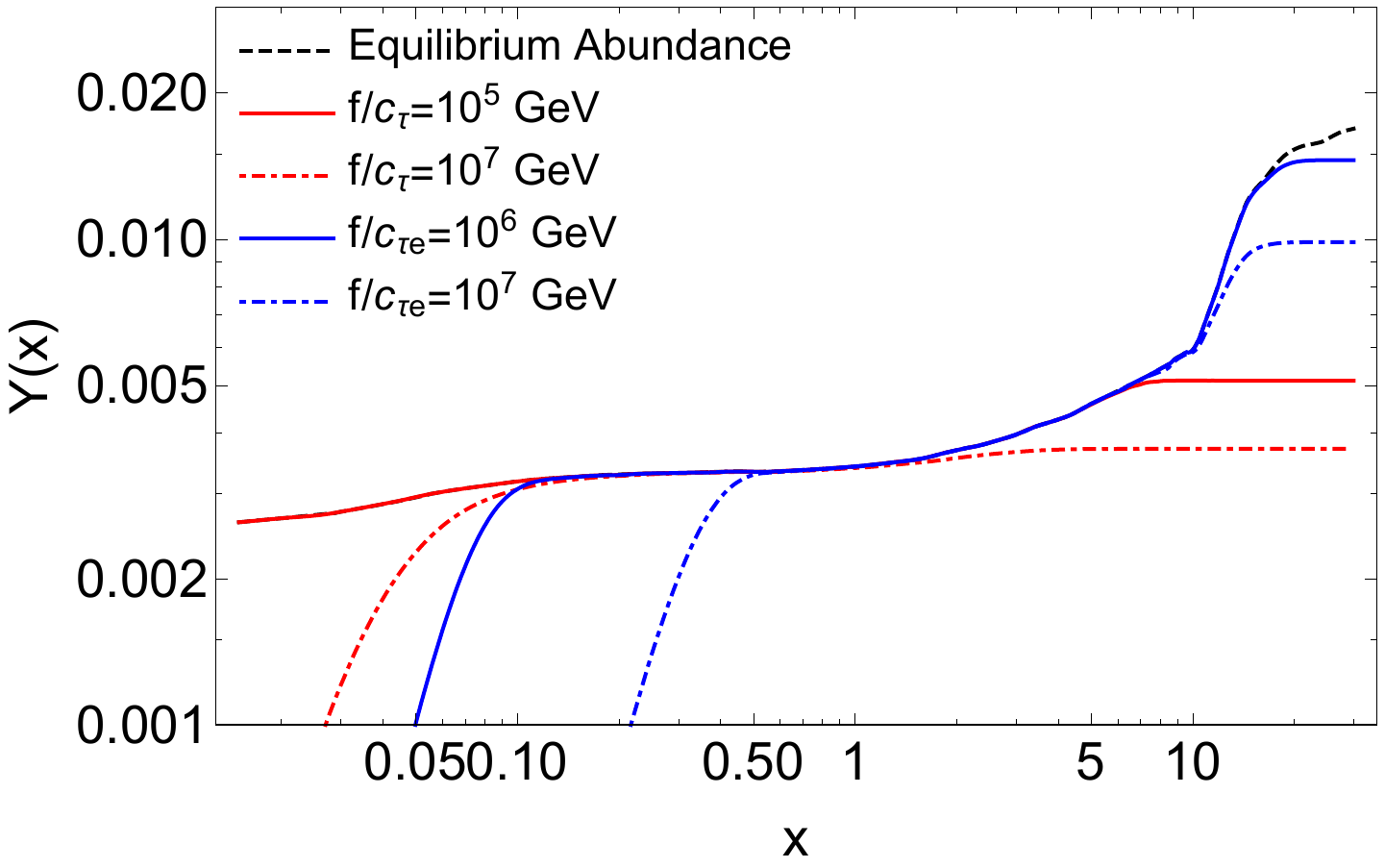}
		\caption {Numerical solutions for an axion coupled to the $\tau$ for both scattering (red) and decay (blue). The dashed black line represents the axion comoving equilibrium number density.}
		\label{fig:Yvsx}
	\end{center}
\end{figure}

\subsection{Numerical Solutions}

We show numerical results in Fig.~\ref{fig:Yvsx} for scattering (red lines) and decay (blue lines) involving the $\tau$ lepton. We choose here the values of $g_*$ and $g_{*s}$ from Ref.~\cite{Borsanyi:2016ksw}. As manifest from the figure, and also as anticipated by the semi-analytical solution in Eq.~\eqref{eq:Yanalytical}, axions are produced mostly at temperatures $T \simeq m_\tau$. This feature is understood upon comparing the temperature dependence of the production rates with the one for the Hubble parameter, $H \simeq T^2 / M_{\rm Pl}$. 

For scatterings, the rate scales as $\overline{\Gamma}_S \propto (m_\ell/f)^2 T$, whereas for decays once one accounts for Lorentz time dilation we find the scaling $\overline{\Gamma}_D \propto \Gamma_{\ell^\pm \, \rightarrow \, \ell^{\prime \pm}  a} (m_\ell /T)$.~\footnote{The scattering scaling may seem counterintuitive since the operators in Eq.~\eqref{eq:Lint} would naturally lead to $\Gamma_S \simeq T^3 /f^2$. However, the axion is a free field unless leptons with opposite chiralities are coupled, which explains the appearance of $m_\ell$ in the rate. Alternatively, one can perform a chiral rotation and work in a basis where the axion does not couple derivatively but with pseudo-Yukawa interactions, where $m_\ell$ is manifest.} The ratio $\overline{\Gamma}_{S,D} / H$ is saturated at small temperatures for both cases, until we get to values around $m_\ell$ and the scatterers number density is Maxwell-Boltzmann suppressed. We also show the equilibrium density (dashed black lines), which is not constant due to the $g_{*s}$ temperature dependence. Larger couplings keep the axion in thermal equilibrium down to lower temperatures, so they result in a larger axion relic density and a larger contribution to $\dN$, which we now quantify. 

\section{Axion Contribution to $\Delta N_{\rm eff}$} 
\label{sec:dN}

The output of the Boltzmann equation is the asymptotic value of $Y_a$ as a function of the axion coupling to leptons. This quantity is directly connected to the number of effective neutrinos, defined from the expression of the radiation energy density in the late universe as follows
\be
\rho_{\rm rad} = \rho_\gamma + \rho_\nu + \rho_a \equiv \left[ 1 + \frac{7}{8} \left( \frac{T_\nu}{T_\gamma} \right)^{4} N_{\rm eff} \right] \rho_\gamma \, . 
\ee 
In the absence of physics beyond the SM, the prediction is $N_\text{eff(SM)} = 3.046$,~\footnote{Neutrino decoupling is not instantaneous, hence the correction to the naive prediction $N_\text{eff(SM)} = 3$~\cite{Dodelson:1992km,Hannestad:1995rs,Dolgov:1997mb,Mangano:2005cc,deSalas:2016ztq}.} whereas the presence of axions results in the deviation
\be
\Delta N_{\rm eff} = N_{\rm eff} - N_\text{eff(SM)} = \frac{8}{7} \left( \frac{T_\gamma}{T_\nu} \right)^{4} \frac{\rho_a}{\rho_\gamma} = \frac{8}{7} \left( \frac{11}{4} \right)^{4/3} \frac{\rho_a}{\rho_\gamma}  \ .
\label{eq:NeffDEF}
\ee
For a relativistic gas of bosons with $g$ internal degrees of freedom we have
\be
\rho_a = \frac{\pi^2}{30} \left(\frac{\pi^2 n_a}{\zeta(3) g}\right)^{4/3} \ ,
\ee
where $\zeta(3) \simeq 1.2$. A similar manipulation relates the photon energy and entropy densities
\be
\rho_\gamma = 2 \times \frac{\pi^2}{30} \left(\frac{45 \,s}{2 \pi^2 g_{*s}}\right)^{4/3} \ .
\ee
Taking the ratio between the last two expressions and plugging the result into Eq.~\eqref{eq:NeffDEF}, we find the axion contribution to the effective number of neutrino species
\be
\Delta N_{\rm eff} \simeq 12.15 \, \left( g_{*s} \, Y_a\right)^{4/3} \simeq 74.85 \, Y_a^{4/3} \ ,
\ee
where we use the late universe value $g_{*s} = 2 + (7/8) \times 2 \times 3 \times (4/11) = 43/11$.

We estimate the contribution to $\dN$ from the semi-analytical solution in Eq.~(\ref{semianalytical}), and we find the approximate expression
\begin{eqnarray}
\Delta N_{\rm eff} \simeq \frac{4}{7} \, \left(\frac{43}{4g_{*}}\right)^{\frac{4}{3}} \left[1-e^{-\frac{\Gamma}{H}\big \rvert_{T=m_\ell}}\right]^\frac{4}{3} \, .
\end{eqnarray}
For large $\Gamma/H|_{T=m_\ell}$, axions reach thermal equilibrium and $\dN$ is then fixed by the value of $g_{*s}$ when they decouple. For small $\Gamma/H|_{T=m_\ell}$, we expand the previous expression
\begin{eqnarray}
\Delta N_{\rm eff} \simeq \frac{4}{7} \, \left(\frac{43}{4g_{*}}\right)^{\frac{4}{3}} 
\left[ \frac{\Gamma}{H}\bigg \rvert_{T=m_\ell} \right]^{4/3} 
\propto \left( \frac{c_\ell}{f} \right)^\frac{8}{3} \, ,
\end{eqnarray}
since all production rates scale as $(c_\ell/f)^{2}$. This last expression is valid in the large $f$ case, when axions do not thermalize. Similarly, one can solve the Boltzmann equation for the case of production via decays, finding the same asymptotic behavior at large $f$.

\begin{figure}
	\begin{center}
		\vspace*{3mm}
		\includegraphics[width=0.8\linewidth]{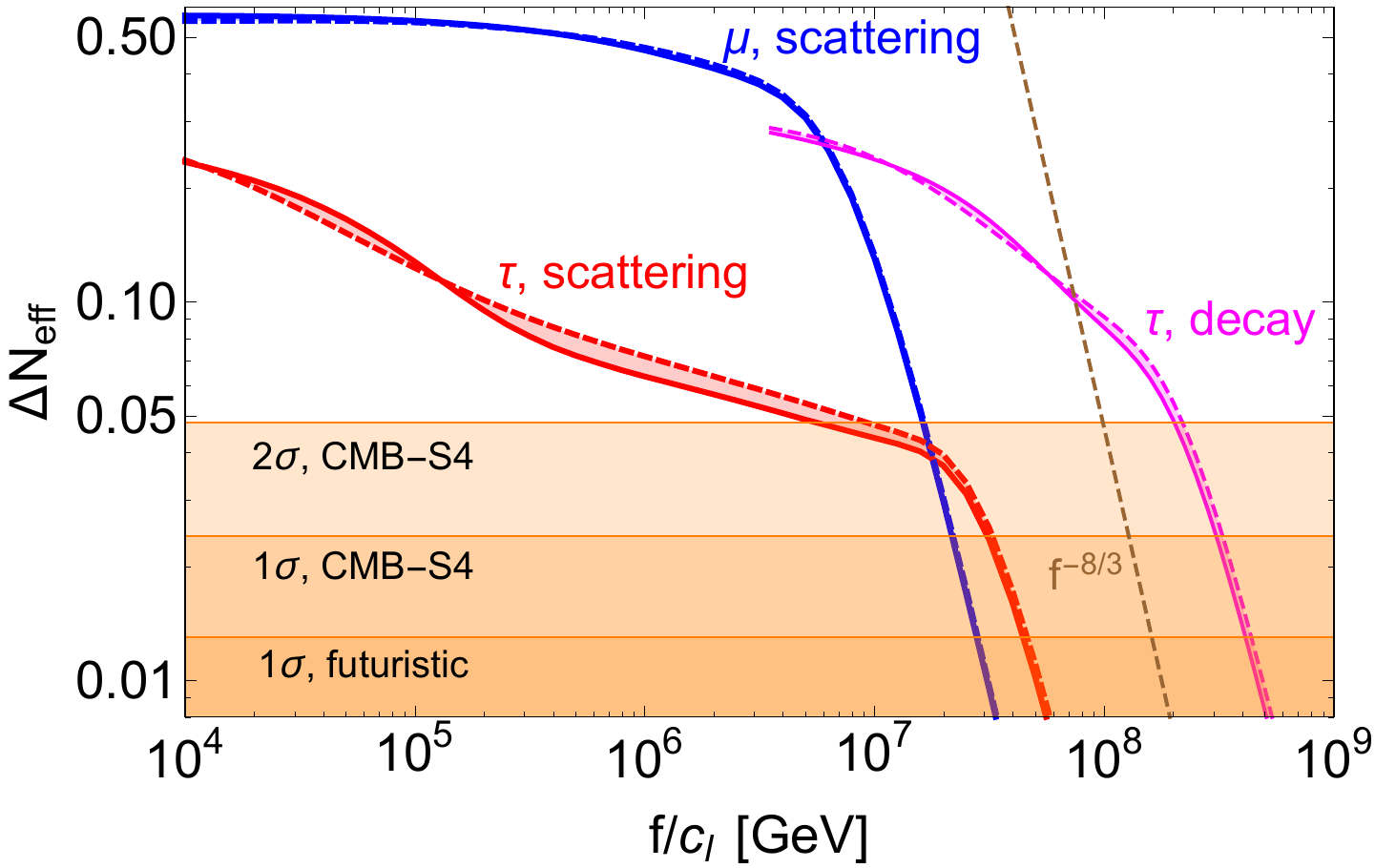}
		\caption {Contribution to $N_{\rm eff}$ from muon (blue) and tau (red) scattering as a function of $c_\ell/f$. Decays are possible for off-diagonal couplings; we show only results for tau decays (magenta), which are the only allowed ones in the above range for $c_{\ell \ell^\prime}/f$. Each process is shown as a band, parameterizing the uncertainty in the number of relativistic degrees of freedom: the straight line $g_*$ is taken from~\cite{Laine:2006cp} and the dashed line from~\cite{Borsanyi:2016ksw}. We also show the analytical expectation, $N_{\rm eff} \propto f^{-8/3}$, for non-thermalized axions. The orange bands represent the forecasted sensitivities for future CMB experiments~\cite{Abazajian:2016yjj}.}
		\label{figura}
	\end{center}
\end{figure}

A  quantitative prediction for $\Delta N_{\rm eff}$ requires solving the Boltzmann equation numerically. We show in Fig.~\ref{figura} the predicted $\Delta N_\text{eff}$ as a function of $f/c_\ell$ for $\ell=\mu,\tau$ and, in the case of tau decays, as a function of $f / c_{\tau,\ell^\prime}$, where $\ell^\prime = {e,\mu}$. Given the hierarchy $m_\tau \gg m_{\mu, e}$, the two decay channels have almost the same kinematics and the associated rates can differ only because of the value of $c_{\tau,\ell^\prime}$. The two curves correspond to the use of two different $g_*$ and $g_{*s}$~\cite{Laine:2006cp,Borsanyi:2016ksw}. We will refer to the upper curve in the rest of the analysis. We observe a change in behavior between the small $f$ regime, where equilibrium is reached, and the large $f$ regime, where equilibrium is never reached. Scatterings with $\tau$ thermalize the axion if $f/c_\tau \lesssim 2\times 10^7$ GeV, leading to $\dN \gtrsim 0.04$ and reaching $\dN \simeq 0.2$ at $f/c_\tau \simeq 10^4$ GeV. In the case of scatterings with muons, thermalization happens for $f/c_\mu\lesssim 3\times 10^6$ GeV giving $\dN \gtrsim 0.33$ and reaching $\dN \simeq 0.57$ at $f/c_\mu \simeq 10^4$ GeV. We take $10^4$ GeV as a benchmark lower bound, given the model dependence of the constraints discussed above. We could, in principle, explore even lower values of those couplings. Finally, for $\tau$ decays, axions thermalize for $f/c_{\tau l^\prime} \lesssim 2\times 10^8$ GeV with $\dN \gtrsim 0.05$ and giving $\dN \simeq 0.28$ at $f/c_{\tau l^\prime} \simeq 3 \times 10^6$ GeV. Lower values of $f/ c_{\tau l^\prime}$ are experimentally excluded. For the same reason, we do not plot results for muon decays since present constraints forbid a large $\dN$ through this process.

\section{The QCD Axion case}
\label{sec:QCD}

We explore then the possibility that the bosonic field $a$ is the well motivated QCD axion. If this is the case, there are several complementary experimental constraints from: (1) a coupling to gluons that induces a mass term; (2) a coupling to photons that arises necessarily from axion-pion mixing, plus model dependent UV contributions; (3) couplings to nucleons; (4) a possible direct coupling to electrons.

The coupling to gluons, as in Eq.~\eqref{eq:Lint} with the conventional normalization $C_{GG} = 1$, induces a periodic potential from QCD non-perturbative effects. Upon expanding such a potential around the minimum, we find a zero-temperature axion mass~\cite{DiVecchia:1980yfw,Georgi:1986df}
\begin{eqnarray}
m_a = 0.57 \left( \frac{10^7 {\rm GeV}}{f} \right) {\rm eV}. \label{relax}
\end{eqnarray}
Comparing with Fig.~\ref{figura}, axions relativistic at matter-radiation equality ($m_a\ll {\cal O}({\rm eV})$) and giving $\dN \approx 0.2$ are consistent with production driven by $\mu$ scatterings or $\tau$ decays with ${\cal O}(1)$ couplings $c_{\mu}$ and $c_{\tau \ell'}$. In order to reach $\dN\approx 0.4$, one would need $c_\mu \simeq 5$. Instead, production via $\tau$ scattering requires $c_\tau \approx {\cal O}(10^2)$ to have such large values of $\dN$. 

The coupling to photons is typically expected to be $C_{\gamma \gamma} \simeq {\cal O}(1)$. The strongest bounds on these interactions come from the cooling of horizontal branch stars~\cite{Raffelt:2006cw,Ayala:2014pea,Patrignani:2016xqp}
\begin{eqnarray}
f/ C_{\gamma \gamma} \gtrsim 10^7 {\rm GeV} \, ,
\end{eqnarray}
and from the CAST experiment~\cite{Arik:2008mq}
\be
f/ C_{\gamma \gamma}  \gtrsim \left\{ \begin{array}{lccccc}
1.8 \times 10^7 \, {\rm GeV} & & $\quad\quad$ & & &  m_a\lesssim 10^{-2} {\rm eV}  \vspace{0.1cm} \\
10^6 \, {\rm GeV}  &  & $\quad\quad$  & & & 10^{-2} {\rm eV} \lesssim m_a\lesssim 0.4 \, {\rm eV}
\end{array} \right. \ .
\ee
For $C_{\gamma \gamma} = {\cal O}(1)$, this implies a range similar to the one obtained by imposing the axions to be relativistic at matter-radiation equality in Eq.~(\ref{relax}). However, cancellations between model dependent and model independent contributions are possible, implying that the above bounds can be relaxed. 

Another important constraint on the QCD axion comes from the SN 1987A cooling process due to coupling with nucleons~\cite{Raffelt:2006cw,Fischer:2016cyd,Giannotti:2017hny,Patrignani:2016xqp}. Such bounds do not apply to an ALP coupled only to leptons, since no interactions with gluons and nucleons are induced at one-loop. The couplings to nucleons for the QCD axion are model dependent, since it also depends on the coupling to quarks; upon allowing cancellations~\cite{DiLuzio:2017ogq} the bounds can be relaxed. Moreover, such bounds are rather uncertain due to complicated SN physics. With these {\it caveats} in mind, we take $f\gtrsim 4\times 10^8$ GeV as a reference value~\cite{Raffelt:2006cw,Patrignani:2016xqp} and we find that a QCD axion can lead to an $N_{\rm eff} \approx 0.2$ only for $c_\mu$ or $c_{\tau \ell'}$ of about 100, or for $c_\tau \approx 10^4$.

\section{Cosmological implications of hot axions}
\label{sec:cosmo}

Hot axions are a microscopic realization for the phenomenological $\Lambda$CDM+$\dN$ model, hence providing a theoretical motivation for $N_{\rm eff} > N_\text{eff(SM)} = 3.046$. In this section, we report the effects of hot axions on the cosmological parameters. Our results are valid either for massless or extremely light axions, as long as their mass does not affect cosmological observables. The tools we employ are the public Monte Carlo Markov chains from Planck Collaboration,\footnote{All the public results from Planck Collaboration regarding cosmological parameter inference can be found in \href{http://pla.esac.esa.int/pla/}{http://pla.esac.esa.int/pla/}.} corresponding to $\Lambda$CDM+$N_{\rm eff}$ (selecting only the posterior distribution within $N_{\rm eff}>3.046$). We resort to the full Planck 2018 temperature, polarization and cross correlation angular power spectra~\cite{Planck2018_cosmopars}, along with BAO measurements from the last data release from BOSS~\cite{Alam:2016hwk} and lower redshift BAO measurements~\cite{Beutler11,Ross15}.

\begin{figure}
	\centering
	\includegraphics[width=0.8\linewidth]{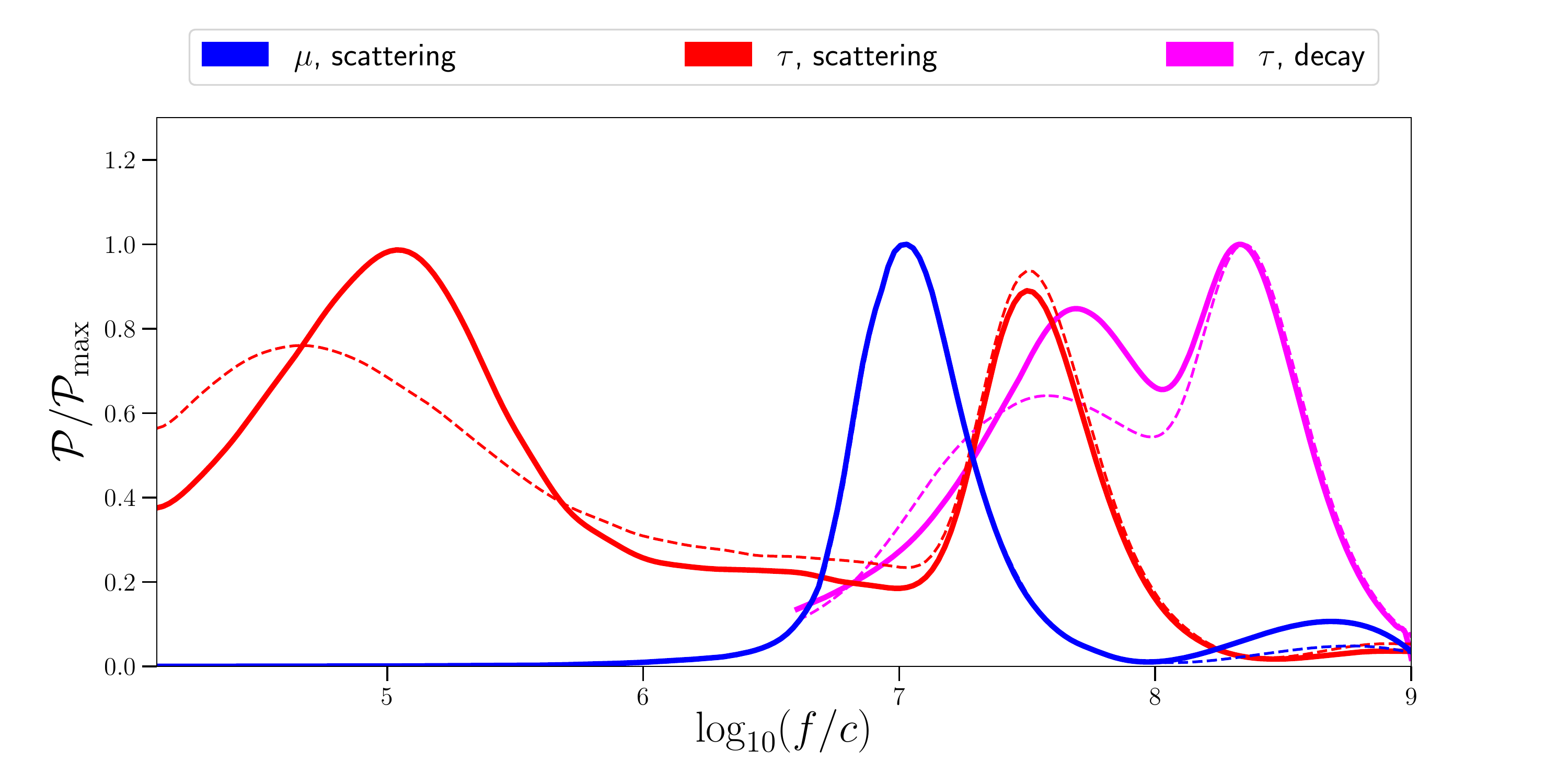}
	\caption{Posterior distribution of $\log_{10}(f/c_\ell)$ obtained assuming a $\Lambda$CDM+$\dN$ model with flat priors in $\dN$ and considering only the values of $\dN$ accessible for each case (see Fig. \ref{figura}). We show results for axion production via $\mu$ scatterings (blue), $\tau$ scattering (red) and $\tau$ decays (magenta). We do not show the parameter space corresponding to $f/c_{\tau \ell^\prime} < 3 \times 10^6 \, {\rm GeV}$ since it is ruled out by the constraints discussed in Sec.~\ref{Lag}. Solid and dashed lines assume $g_*$ from Refs.~\cite{Laine:2006cp} and~\cite{Borsanyi:2016ksw}, respectively. The full Planck angular power spectra and BAO data were used in all cases.
	}
	\label{fig:post_f}
\end{figure}

As a first step of our analysis, we assume flat priors on $N_{\rm eff}$ in the range $3.046 \leq N_{\rm eff}\leq \infty$ as done in Ref.~\cite{Planck2018_cosmopars}, or equivalently $0 \leq \dN \leq \infty$. Considering the range of $\dN$ accessible for each case in Fig.~\ref{figura}, we derive the corresponding marginalized posterior distribution for $\log (f/c_\ell)$ shown in Fig.~\ref{fig:post_f}. We give results for both $g_*$ obtained by Refs.~\cite{Laine:2006cp} and~\cite{Borsanyi:2016ksw} shown by solid and dashed lines, respectively. The change of variable needed to obtain the posterior distribution requires the Jacobian factor ${\rm d}\dN/{\rm d}\log_{10}(f/c_\ell)$, and as emphasized in Sec.~\ref{Sec:production} axions are mostly produced around the QCDPT. As a result, the posterior distribution of $\log_{10}(f/c_\ell)$ has small changes for different choices of $g_*$. Within this theoretical uncertainty, we conclude from Fig.~\ref{fig:post_f} that axions produced via $\mu$ scattering indicate $f/c_\mu\sim 10^7$ GeV, while if the production is driven by $\tau$ scattering we have $f/c_\tau\sim 10^{4.5}$-$10^5$ GeV or $10^{7.5}$ GeV. Finally, production via $\tau$ decays implies $f/c_\tau$ between $10^{7.5}$ GeV and $10^{8.5}$ GeV.

However, one should note that a flat prior on $\dN$ disfavors large values of $f/c_{\ell}$; the whole interval $10^9 \,{\rm GeV}\lesssim f/c_{\ell}\lesssim 10^{18}\,{\rm GeV}$ predicts $\dN\sim 0$ (see Fig.~\ref{figura}). As we have shown in the previous section, $\dN$ is intimately related with the different couplings $f/c_\ell$. Therefore, we also perform the analysis starting from this theoretical prior knowledge rather than from a phenomenological model. We assume flat priors on $\log_{10}(f/c_{\ell})$ on the range $(f/c_{\ell})|_{\rm min} < f/c_{\ell} < (f/c_{\ell})|_{\rm max}$, then we derive the corresponding priors on $\dN$ and finally we analyze the Planck chains applying these priors by importance sampling. The value $(f/c_{\ell})|_{\rm min}$ is determined by the experimental constraints, which, as we mentioned before, have some model dependences. The choice for $(f/c_{\ell})|_{\rm max}$ is more arbitrary. On purely theoretical grounds one could argue that the maximal scale is probably the Planck scale. Another sensible choice could be around $10^{11}$ GeV. Indeed, the QCD axion would give too much DM for larger values, unless PQ is broken before inflation and not restored afterwards with a finely-tuned initial misalignment angle~\cite{Bae:2008ue}. This discussion may seem purely academic, because the entire parameter range $f/c_{\ell} \gg 10^8 - 10^9$ GeV gives a negligible effect on $\dN$. However, a very large $(f/c_{\ell})|_{\rm max}$ combined with a flat prior on $\log_{10} (f/c_{\ell})$ implies a large peak on the prior probability at $\dN\rightarrow 0$, which may artificially drive the posterior distribution towards such values. In order to show the dependence on such an assumption we consider the three different priors $(f/c_{\ell})|_{\rm max} = \left(10^{18}, 10^{11}, 10^8\right)$ GeV. The latter value corresponds to the maximum $f/c_{\ell}$ for which $\dN$ is significantly different from zero, with the present experimental sensitivity. We refer the interested reader to App.~\ref{App:Cosmo} for more details on the priors on $\dN$. 

\begin{figure}
	\centering
	\includegraphics[width=\linewidth]{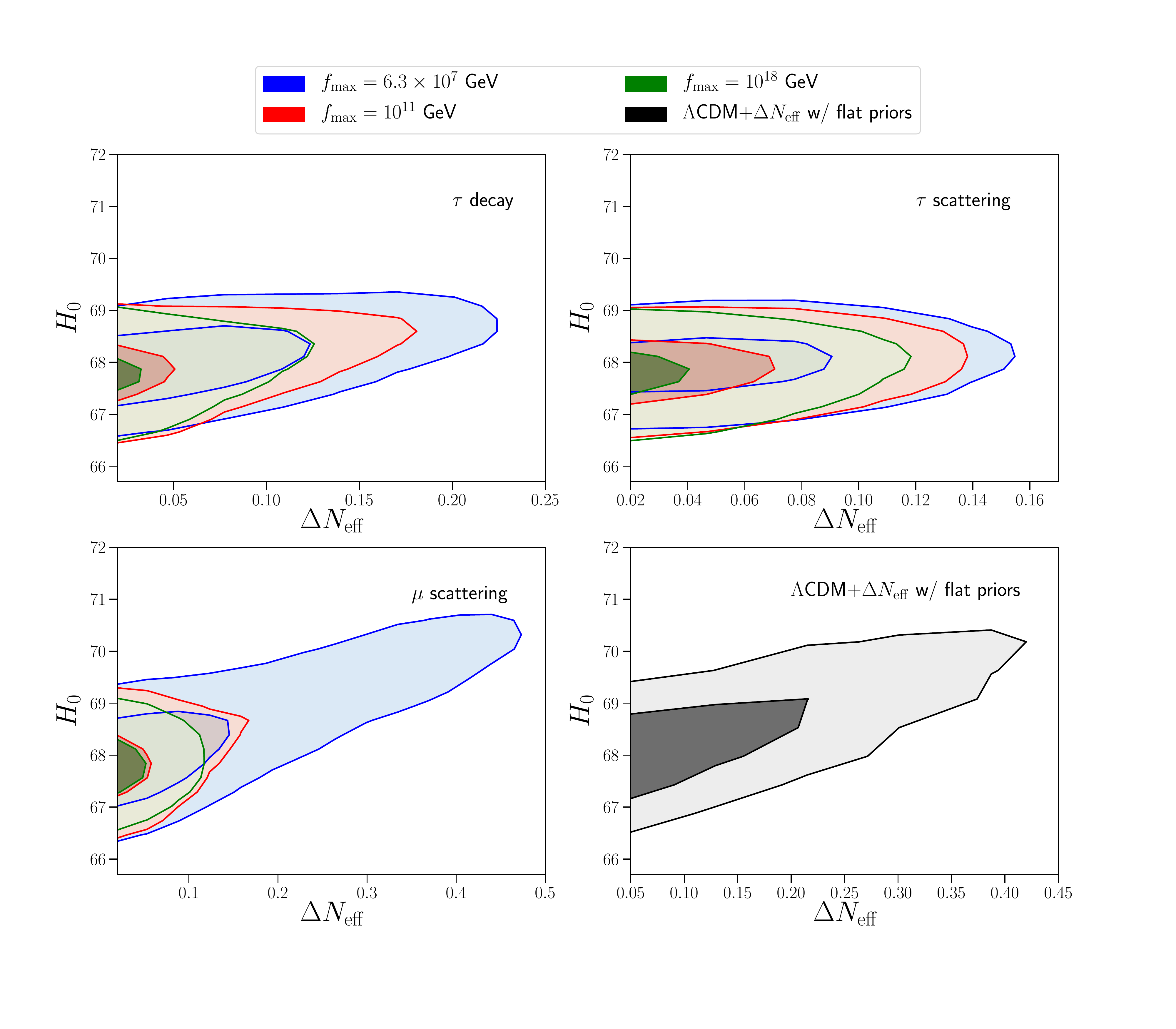}
	\caption{68\% and 95\% confidence level regions in the $(H_0, \dN)$ plane for hot axions assuming axion production via decays and scattering with tau (upper left and upper right panels, respectively) and scattering with muons (lower left panel). We consider three different priors (shown in different colors), all of them flat in $\log_{10}(f/c_{\ell})$  but with different maximum values of $f/c_{\ell}$, as indicated in the legend. We also show the constraints (bottom right panel) assuming a standard $\Lambda$CDM+$\dN$ model with flat priors in $\dN$ > 0 . The full Planck angular power spectra and BAO data were used in all cases. Note the change of scale in the horizontal axis in each panel.
		\label{fig:H0-dN}}
\end{figure}

The marginalized constraints on the plane $(\dN, H_0)$ are shown in Fig.~\ref{fig:H0-dN} for the three possible channels of interaction between leptons and the hot axions and considering the three priors. Given that the priors are bimodal ({\it i.e.}, they favor either $\dN\sim 0$ or $\dN\sim 0.15$-$0.4$, depending on the case), we obtain large tails towards large values of $\dN$. However, the maximum values of $\dN$ are limited for each coupling, and the peak at large $\dN$ in the priors does not dominate over the peak at $\dN\rightarrow 0$. Then, the resulting $H_0$ values are lower than those obtained assuming standard $\Lambda$CDM+$\dN$, shown in the lower right panel of Fig.~\ref{fig:H0-dN}. The only exception appears when the coupling with muons via scattering is assumed and the prior with the lowest upper bound for $f$ is chosen, whose results are similar to the standard $\Lambda$CDM+$\dN$. Nonetheless, it is worth to emphasize that hot axions provide a motivation for $\dN$, and that the results are qualitatively similar to those coming from a non motivated theoretical framework and follow, instead, a phenomenological parametrization.

We also evaluate the tension of the marginalized posterior distribution of $H_0$ from our results with respect to the direct measurement of $H_0$. We assume a Gaussian posterior distribution for the direct measurement with mean 73.52 km~s$^{-1}$~Mpc$^{-1}$ and width 1.62~km~s$^{-1}$~Mpc$^{-1}$~\cite{Riess:2018byc}. To compute the tension, $\mathcal{T}$, between two experiments which have measured values $A$ and $B$:
\begin{equation}
\mathcal{T}=\frac{\lvert A-B\lvert}{\sqrt{\sigma_A^2+\sigma_B^2}}\, , \qquad\qquad \mathcal{T}^*=2\frac{\lvert A-B\lvert}{\sqrt{(\sigma_A^*)^2+(\sigma_B^*)^2}}\, .
\label{eq:tension}
\end{equation} 
In these expressions, $\sigma_i$ and $\sigma^*_i$ correspond to the 68\% and 95\%  confidence level error of the measurement $i$, respectively. Then, $\mathcal{T}$ can be interpreted as the distance in 68\% errors between the two measurements. $\mathcal{T}^*$ is the equivalent of $\mathcal{T}$ computed using the 95\% confidence level errors instead, which is useful when the distribution is not Gaussian and $\sigma^*_i \neq 2\sigma_i$. Thus, for Gaussian posteriors, $\mathcal{T}=\mathcal{T}^*$.

\begin{table}[]
\centering
\begin{tabular}{|c|c|c|c|c|}
\hline
Model & Coupling & Prior $(f/c)_{\rm max}$ [GeV]& $H_0$ {[}${\rm km\, s^{-1}\, Mpc^{-1}}${]} & $\mathcal{T}$ ($N\sigma$) \\ \hline
\multirow{10}{*}{$\Lambda$CDM+$\dN$} & \multirow{3}{*}{$\mu$ scattering} & $3\times 10^7$  & $68.0 ^{+0.8 } _{ -0.7 } (^{ +2.3 } _{ -1.1 })$ & 3.06 (2.75$^*$) \\
 &  & $10^{11}$  & $67.8 ^{+0.6 } _{ -0.5 } (^{+1.4 } _{ -1.1 })$ & 3.36 \\
 &  & $10^{18}$  & $67.7 ^{+0.5 } _{ -0.4 } (^{+1.2 } _{ -1.0 })$ & 3.38 \\ \cline{2-5} 
 & \multirow{3}{*}{$\tau$ decay} & $6.3\times 10^7$ GeV & $68.1 ^{+0.6 } _{ -0.5 } (^{+1.2 } _{ -1.0 })$ & 3.18 \\
 &  & $10^{11}$  & $67.8 ^{+0.6 } _{ -0.5 } (^{+1.2 } _{ -0.9 })$ & 3.35 \\
 &  & $10^{18}$  & $67.7 ^{+0.5 } _{ -0.4 } (^{+1.1 } _{ -0.9 })$ & 3.39 \\ \cline{2-5} 
 & \multirow{3}{*}{$\tau$ scattering} & $5\times 10^8$ & $68.0 ^{+0.5 } _{ -0.5 } (^{+1.0 } _{ -1.0 })$ & 3.25 \\
 &  & $10^{11}$  & $67.8 ^{+0.5 } _{ -0.5 } (^{+1.1 } _{ -1.0 })$ & 3.33 \\
 &  & $10^{18}$  & $67.7 ^{+0.5 } _{ -0.5 } (^{+1.1 } _{ -0.9 })$ & 3.39 \\ \cline{2-5} 
 & No coupling & - & $68.3 ^{+0.8 } _{ -0.7 } (^{+1.8 } _{ -1.2 })$ & 2.93 \\ \hline
$\Lambda$CDM+$N_{\rm eff}$ & No coupling & - & $67.4 ^{+1.1 } _{ -1.2 } (^{ +2.3 } _{ -2.3 })$ & 3.08 \\ \hline
$\Lambda$CDM & No coupling & - & $67.7 ^{+0.5 } _{ -0.4 } (^{+0.9 } _{ -0.9 })$ & 3.46 \\ \hline
\end{tabular}
\caption{
Highest marginalised posteriors density values of $H_0$ and $68\% $ ($ 95\% $ in parenthesis) highest density intervals,
for each model, coupling and prior choice. The models with ``No coupling'' correspond to the standard phenomenological parametrization with flat priors in $\dN$. $\Lambda$CDM+$ N_{\rm eff}$ refers to the same model as $\Lambda$CDM+$\dN$, but without imposing $N_{\rm eff}>3.046$. In the right-most column, we show the tension with respect to the local measurement of $H_0$, using the first expression in Eq.~\ref{eq:tension}. $^*$ The posterior distribution of $H_0$ in this case has a large tail towards higher $H_0$ (see main text and fig.~\ref{fig:H0-dN}); if the tension is computed using the second expression in Eq.~\ref{eq:tension}, the resulting tension is 2.75.
}
\label{tab:tension}
\end{table}

The marginalised constraints on $H_0$ obtained for each coupling and prior choice are shown in Tab~\ref{tab:tension}, along with the tension for the standard $\Lambda$CDM+$N_{\rm eff}$ (where $\dN<0$ is allowed) and $\Lambda$CDM for comparison. In all cases, the full Planck 2018 angular power spectra plus BAO data are used. As can be seen, the maximum value of the marginalised posteriors and the uncertainties do not vary significantly for most cases. However there are two exceptions. On one hand, the errors on $H_0$ for the hot axions coupled with the muons and using the lowest upper bound of the prior on $\log_{10}(f/c_\ell)$ are larger than in the rest of the cases, especially towards higher values, because the constraints on $\dN$ are weaker in this case. The large tail towards higher $H_0$ supposes also that the tension reported obtained using the first expression in Eq.~\ref{eq:tension}, $\mathcal{T}$, is overestimated.  The tension correctly computed in this case is 2.75, corresponding to the second expression in Eq.~\ref{eq:tension}, $\mathcal{T}^*$.  On the other hand, in the $\Lambda$CDM case with flat prior on $\dN$, the constraints on $\dN$ and $H_0$ are again weaker. Note also that if $N_{\rm eff}>3.046$ is not imposed, the errors on $H_0$ increase significantly, but the central value is shifted towards lower $H_0$. 

A positive contribution to the number of effective neutrinos is one of the preferred solutions to alleviate the tension between the different measurements of the Hubble constant. Although axions provide a motivated realization for $\dN\neq 0$, the tension is generically only slightly reduced with respect to the one found assuming $\Lambda$CDM. Nonetheless, the tension decreases from 3.46 to 2.75 when one assumes that the axions are coupled to the muons via scattering and the prior on $\log_{10}(f/c)$ with the lowest upper bound is used. 

\section{Conclusion} 
\label{sec:Conclusion}

We have studied hot axions production through scatterings and decays of heavy leptons (muon and tau). Our results hold for a generic ALP as well as for the QCD axion. Axion production via fermion scatterings was originally proposed in Ref.~\cite{Turner:1986tb}, previous estimates of $\dN$ due to leptons were given in Refs.~\cite{Brust:2013xpv,Baumann:2016wac}. Here, we have computed cross sections and decay rates, we have evaluated the thermal averages of these quantities to find the temperature dependent production rates and we have numerically solved the Boltzmann equation to compute the axion abundance. As a final result, we have found the axion contribution to the effective number of additional neutrinos  $\dN$ as a function of the axion couplings. As summarized in Fig.~\ref{figura}, couplings within the allowed parameter space can lead to large signals: $\dN^\text{max} \simeq 0.6$ for muon scatterings, $\dN^\text{max} \simeq 0.3$ for tau decays and $\dN^\text{max} \simeq 0.2 $ for tau scatterings. These scenarios provide a well motivated particle physics framework for $\dN >0$ that can be further explored by current and future experiments~\cite{SPT-3G,Abazajian:2016yjj}. 

Given such results, we have also investigated the consequences for the current $3.6\sigma$ ($3.46\sigma$, combining with BAO) tension between high and low redshift measurements of the Hubble constant $H_0$. As already explored, {\it e.g.}, in \cite{Bernal:2016gxb}, large values of $\dN$ can alleviate the tension between the two datasets. Here, we have performed a similar analysis but with theory-based priors by choosing a flat prior in $\log(f/c_\ell)$, with the axion-lepton effective coupling defined as $c_\ell/f$. Our findings feature some dependence on the maximum value considered in the prior distribution, $(f/c_\ell)_{\rm max}$, since the probability of having very small $\dN$ is enhanced at large $(f/c_\ell)_{\rm max}$. Values $\dN \gtrsim 0.2$ are disfavored by the latest Planck 2018 temperature and polarization data combined with BAO data, so the tension cannot be completely erased by a non-vanishing $\dN$. In the scenarios studied here and summarized in Tab.~\ref{tab:tension}, we have also generically found moderate improvement in mitigating the $H_0$ tension. A notable exception is for hot axions produced via muon scattering and $(f/c_\ell)_{\rm max} \simeq 3\times 10^7$, where the tension is reduced to $2.75 \sigma$. Our findings provide a theoretically motivated origin for $\dN >0$ and motivate further studies of UV complete models with axions coupled to heavy leptons. Hot axions produced around or below the QCDPT open an exciting window to observe extremely weakly-coupled pseudo-scalars. Forthcoming results from CMB surveys and direct searches make the future of these hypothetical dark components of our universe very bright.

\acknowledgments
We thank Fernando Arias Aragon, Jorge de Blas, Massimilano Lattanzi and Miguel Quartin for useful discussions. F.D. was supported by Istituto Nazionale di Fisica Nucleare (INFN) through the ``Theoretical Astroparticle Physics'' (TAsP) project. A.N. and R.Z.F. were supported by the grants EC FPA2010-20807-C02-02, AGAUR 2009-SGR-168. J.L.B. is supported by the
Spanish MINECO under grant BES-2015-071307, co-funded by the ESF.  A.N., R.Z.F and J.L.B. were supported by  the Spanish MINECO under MDM-2014-0369 of ICCUB (Unidad de Excelencia ``Maria de Maeztu''). F.D. thanks the Institute of High Energy Physics of the Chinese Academy of Sciences, where part of this work was carried out, for hosting him during a secondment of the InvisiblesPlus RISE project. A.N. is grateful  to the Physics Department of the University of Padova for the hospitality and was supported by the ``Visiting Scientist" program of the University of Padova.
\appendix

\section{SN 1987A bound on the axion-muon coupling}
\label{App:Axion-muon}

In this appendix we reassess the bound on the coupling $f/c_\mu$ derived in Ref.~\cite{Brust:2013xpv}, which relied on the observation of the neutrino burst from the SN 1987A. The presence of axions coupled to SM leptons would imply a shorter duration for the burst, with a consequent bound on the rate for energy loss per unit mass  into axions~\cite{Raffelt:2006cw}
\begin{eqnarray} \label{energy constraint}
\dot{\epsilon}_a < 10^{19} \, \rm{erg}\, s^{-1} g^{-1}\,,
\end{eqnarray}
for a mass density $\rho \simeq 3\times 10^{14}$ g $\text{cm}^{-3}$ and temperature estimated to be around $T \simeq 20-60$ MeV. The energy lost into axions coupled to a SM lepton $\psi$ is given by~\cite{Fukugita:1982gn}
\begin{eqnarray}
\dot{\epsilon}_a = \zeta[6] \frac{n_\psi}{\rho} \frac{40 \, \alpha_{\rm em}}{\pi^2}  \left(\frac{m_\psi}{f}\right)^2 \frac{T^6}{m_\psi^4}\,.
\end{eqnarray}
For the case of an electron the best bound comes from red giants and it is given in Eq.~(\ref{c_e}). For muons the best bound from stellar physics comes from the SN 1987A. However, as already emphasized in this paper, the muon abundance is highly uncertain. From now on, we assume a thermal spectrum with negligible chemical potential
\begin{eqnarray}
n_\mu = 4\times \frac{m_\mu^2 T}{2\pi^2} K_2 \,\left[m_\mu/T\right]\, .
\end{eqnarray}

The SN temperature, even if not known with high precision, is definitely below the muon mass. Upon expanding the modified Bessel function in this regime, which leads to a Maxwell-Boltzmann muon number density $n_\mu \sim (m_\mu T)^{3/2} \exp[- m_\mu/T]$, we notice how the final result is exponentially sensitive on the exact SN temperature. 

\begin{figure}
	\begin{center}		
		\includegraphics[width=0.8\linewidth]{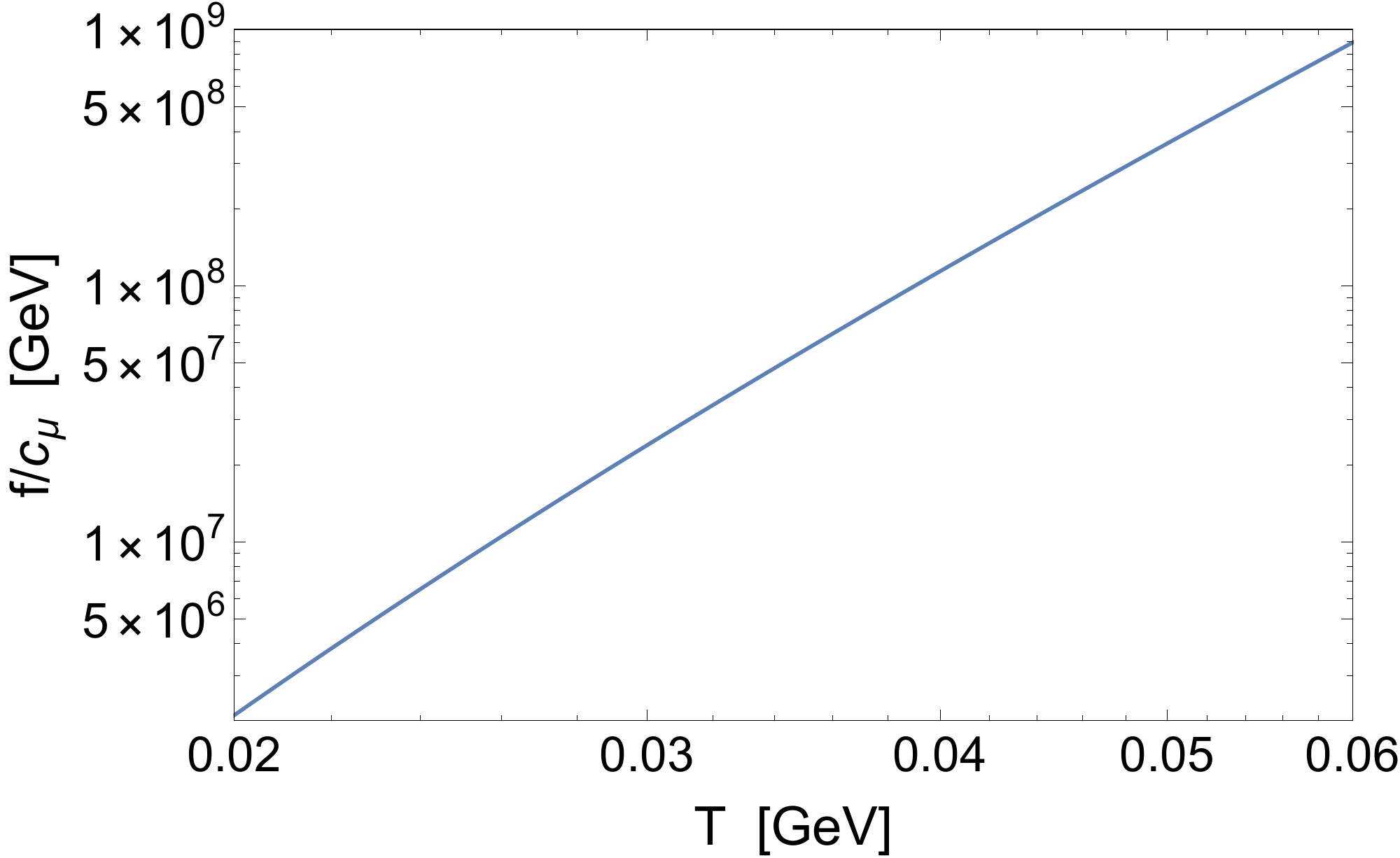}
		\caption{Constraint on $f/c_\mu$ as a function of the SN temperature.	\label{muconst}}
	\end{center}
\end{figure}

 The resulting constraint in Fig.~\ref{muconst} on $f/c_\mu$ shows the strong dependence on the precise value of $T$; as we vary the temperature in the range $T= \left(20,60\right)$ MeV, the constraint varies from $f/c_\mu >10^6$ to $f/c_\mu > 10^9$ GeV. Moreover, as we already mentioned in Sec.~\ref{Sec:constr} there is a big uncertainty on the initial muon abundance. Recent numerical studies~\cite{Bollig:2017lki} have also pointed out that there could be a large muon chemical potential, so more precise calculations are needed in order to have a solid constraint on $f/c_\mu$ from stellar physics.

\section{Radiative corrections to axion-lepton couplings}
\label{app:loop}

In this Appendix, we compute radiative corrections to axion couplings responsible for the bounds in Eqs.~\eqref{eq:loopboundtau} and \eqref{eq:loopboundmu}. Only SM neutral currents appear in the loop diagrams computed here; for this reason, only axion couplings to flavor diagonal currents are affected. 

\begin{figure}
\begin{center}		
\includegraphics[width=0.7\linewidth]{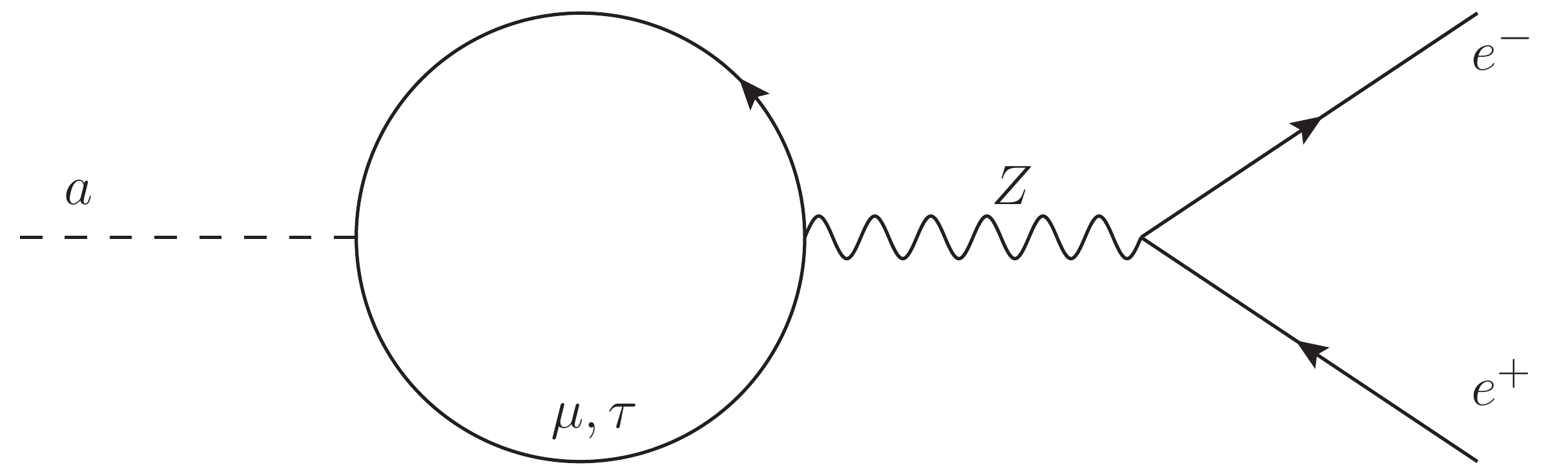}
\caption{Loop diagram inducing an axion coupling to electrons from a coupling to muons or taus.}
\label{fig:Loop}
\end{center}
\end{figure}

We start from the following basis of operators
\be
\mathcal{L}_a^{\rm EFT} = \frac{1}{2f}  \, \partial_\mu a \, \left[ c^0_e \, \bar{e} \gamma^\mu \gamma^5 e + c^0_\mu \, \bar{\mu} \gamma^\mu \gamma^5 \mu + c^0_\tau \, \bar{\tau} \gamma^\mu \gamma^5 \tau \right] \ .
\label{eq:EFTapp}
\ee
The reason for the superscript $0$ is that the coefficients appearing in the above equations are bare couplings before renormalization. We assume these interactions to be generated at the cutoff scale $\Lambda$, which is of the order of the scale $f$ suppressing the operators. Our goal here is to connect the UV values of the Wilson coefficients $c_\ell$ with the correspondent IR quantities, which are the ones relevant for the experimental bounds. Given the wide scale separation between the UV cutoff scale $\Lambda$ and the IR scale (the lepton masses $m_\ell$, where the loop correction is saturated), we only keep the leading logarithmic contribution proportional to $\log(\Lambda / m_\ell)$ from the amplitudes and we neglect finite terms. As shown by the one-loop Feynman diagram in Fig.~\ref{fig:Loop}, an axion coupled to heavy leptons also couples to electrons once radiative corrections are taken into account~\cite{Feng:1997tn}. The amplitude for this diagram is infinite and we regularize the UV divergence by computing the loop integral in $d = 4 - 2 \epsilon$ dimensions.

In order to perform the calculation of the diagram in Fig.~\ref{fig:Loop}, we need to recall the weak gauge interactions between leptons and the Z boson
\be
\mathcal{L}_{\rm NC - SM}^Z = \frac{g}{2 \, c_w}  \, Z_\mu \,
\sum_f   \overline{f}  \gamma^\mu \left(g_V^f + g_A^f \gamma^5  \right) f \ ,
\label{eq:SMNCZ}
\ee
where $g$ and $c_w$ are the $SU(2)_L$ gauge coupling and (the cosine of) the weak mixing angle, respectively. The expression is valid for any SM fermion $f$, with vector and axial-vector coupling to the $Z$ boson given by the relations $g_V^f = T_f^3 - 2 s_w^2 Q_f$ and $g_A^f =  - T_f^3$. Here, $T_f^3$ is the value of the third component of the weak isospin, $s_w$ is the sine of the weak mixing angle an $Q_f$ is the fermion electric charge. 

The loop diagram in Fig.~\ref{fig:Loop} generates a (divergent) correction to the Wilson coefficients $c_\ell$ appearing in Eq.~\eqref{eq:EFTapp}. We are only interested in the correction to $c_e$, so we will focus on this case. In order to properly subtract the divergence, we first identify the tree-level amplitude for the three-point amplitude with external axion, electron and positron
\be
i \, \mathcal{A}_{\rm tree} = \frac{c^0_e}{2 f} \, k_\mu \, \bar{u}(p_1) \gamma^\mu \gamma^5 v(p_2) \ ,
\label{eq:Atree}
\ee
where $k_\mu$ is the axion four-momentum and $u$ and $v$ are the Dirac spinors (also functions of the associated four-momentum). At one-loop, we get a contribution that results in
\be
\begin{split}
i \, \mathcal{A}_{\rm loop} = - \frac{c^0_\ell}{2 f} \, k_\mu &\, \int \frac{d^d q}{(2 \pi)^d} 
 \, {\rm Tr} \left[ \gamma^\mu \gamma^5 \, \frac{i \left( \slashed{q} + \slashed{k} + m_l\right)}{(q+k)^2 - m_\ell^2} \, \frac{i \, g}{2 \, c_w} 
\gamma^\nu \left(g_V^\ell + g_A^\ell \gamma^5  \right) \,  \frac{i \left( \slashed{q} + m_l\right)}{q^2 - m_\ell^2}  \right] \, \times \, \\ &
\frac{- i}{k^2 - m_Z^2} \left(g_{\nu\rho} - \frac{k_\nu k_\rho}{m_Z^2} \right) \times 
\frac{i \,g}{2 \, c_w} \bar{u}(p_1) \gamma^\rho \left(g_V^e + g_A^e \gamma^5  \right)  v(p_2) \ ,
\end{split}
\ee
where the overall minus sign accounts for the fermion loop. It is convenient to rewrite the above one-loop amplitude in a compact form
\be
i \, \mathcal{A}_{\rm loop} =  i \, \frac{g^2}{4 \, c^2_w} \frac{c^0_\ell}{2 f} \,  \,   
\frac{k_\mu}{k^2 - m_Z^2} \left( g_{\nu\rho} - \frac{k_\nu k_\rho}{m_Z^2} \right) \; L^{\mu\nu}(k, m_\ell)  \; \bar{u}(p_1) \gamma^\rho \left(g_V^e + g_A^e \gamma^5  \right)  v(p_2) \ ,
\label{eq:loopamp}
\ee
where we define the loop function
\be
L^{\mu\nu}(k, m_\ell) \equiv \int \frac{d^d q}{(2 \pi)^d} 
 \, \frac{{\rm Tr} \left[ \gamma^\mu \gamma^5 \, \left( \slashed{q} + \slashed{k} + m_l\right) \, \gamma^\nu \left(g_V^\ell + g_A^\ell \gamma^5  \right) \,  \left( \slashed{q} + m_l\right)  \right]}{\left[ (q+k)^2 - m_\ell^2 \right] \left[q^2 - m_\ell^2 \right]}  \ .
 \label{eq:Lmunu}
\ee

The explicit evaluation of such a loop function is our next task. First, we take care of the denominators by using the Feynman parameters
\be
\frac{1}{\left[ (q+k)^2 - m_\ell^2 \right] \left[q^2 - m_\ell^2 \right]} =  \int_0^1 \frac{dx}{\left(l^2 - \Delta\right)^2} \ , \qquad \qquad \qquad \left\{ \begin{array}{ll} l^\mu = & \, q^\mu + k^\mu x \ , \\
\Delta = & \, m_\ell^2 - k^2 x(1-x)
\end{array}  \right. \ .
\ee
We perform this change of variable in the definition of the loop function given in Eq.~\eqref{eq:Lmunu}. Terms in the integrand that are linear in $l$ vanish by parity. Neglecting these pieces, as well as contributions whose trace over the Dirac indices is vanishing, we find
\be
\begin{split}
L^{\mu\nu}(k, m_\ell) = & \,  g_A^\ell \,  \int_0^1 dx \int \frac{d^d l}{(2 \pi)^d} 
 \, \frac{{\rm Tr} \left[ \gamma^\mu \gamma^5 \slashed{l} \gamma^\nu \gamma^5  \slashed{l}  - 
 x(1-x) \gamma^\mu \gamma^5 \slashed{k} \gamma^\nu \gamma^5  \slashed{k} + m_\ell^2 \gamma^\mu \gamma^5 \gamma^\nu \gamma^5 \right] }{(l^2 - \Delta)^2} = \\ &
4 g_A^\ell \,  \int_0^1 dx \int \frac{d^d l}{(2 \pi)^d} 
 \, \frac{2 l^\mu l^\nu - g^{\mu\nu} l^2 - 2 x(1-x) k^\mu k^\nu - \Delta g^{\mu\nu}}{(l^2 - \Delta)^2}  \ ,
\end{split}
\ee
where in the second equality we evaluated the Dirac trace. We now compute the loop integrals. As argued above, we only need to keep the divergent part of the diagram ({\it i.e.}, the $1/\epsilon$ poles) and not the finite pieces. The relevant contributions are given by the expressions~\cite{Peskin:1995ev}
\be
\begin{split}
& \, \int \frac{d^d l}{(2\pi)^d} \frac{1}{(l^2 - \Delta)^2} = \frac{(- 1)^2}{(4 \pi)^{d/2}} \, i \, \frac{\Gamma(2 - d/2)}{\Gamma(2)} \frac{1}{\Delta^{2-d/2}}  = \frac{i}{16 \pi^2 \epsilon} + \ldots \ , \\
&\, \int \frac{d^d l}{(2\pi)^d} \frac{l^2}{(l^2 - \Delta)^2} = \frac{(- 1)}{(4 \pi)^{d/2}} \frac{d}{2} \frac{\Gamma(1-d/2)}{\Gamma(2)} \, \frac{1}{\Delta^{1-d/2}}   = \frac{i}{16 \pi^2 \epsilon} \, 2 \Delta + \ldots \ , \\
& \, \int \frac{d^d l}{(2\pi)^d} \frac{l^\mu l^\nu}{(l^2 - \Delta)^2} = \frac{(- 1)}{(4 \pi)^{d/2}} \, i \, \frac{g^{\mu\nu}}{2} \, \frac{\Gamma(1 - d/2)}{\Gamma(2)} \frac{1}{\Delta^{1-d/2}}  = \frac{i}{16 \pi^2 \epsilon} \frac{\Delta}{2} g^{\mu\nu} + \ldots  \ ,
\end{split}
\ee
and they allow us to identify the $1/\epsilon$ poles in the loop function
\be
L^{\mu\nu}(k, m_\ell) = - \frac{i \, 8 g_A^\ell }{16 \pi^2 \epsilon} \int_0^1 dx  \left[ \Delta g^{\mu\nu} + x (1-x) k^\mu k^\nu \right] \ .
\ee

We go back to the one-loop amplitude and we plug into Eq.~\eqref{eq:loopamp}  the loop function we have just computed. Before doing that, we observe that the axion is on an external leg, so we can impose the on-shell condition $k^\mu k_\mu = k^2 = m_a^2$. Within our framework, the axion is always assumed to be much lighter than the $Z$ boson and any charged lepton, and this allows us to ignore the term proportional to $k^\mu k^\nu$ in the loop function. Ignoring the axion mass also in the other part of the loop amplitude, and identifying the ratio between the masses of the lepton $l$ and the Z boson $m_\ell^2 / m_Z^2 = 2 \lambda_\ell^2 c_w^2 / g^2$ with $\lambda_\ell$ the lepton Yukawa coupling, we find 
\be
i \, \mathcal{A}_{\rm loop} =  - \frac{c^0_\ell}{2 f} \;
\lambda_\ell^2 \; \left(  \frac{4 g_A^\ell g_A^e}{16 \pi^2 \epsilon}  \right) \; k_\mu \; \bar{u}(p_1) \gamma^\mu  \gamma^5 v(p_2) \ ,
\label{eq:loopamp3}
\ee

The full amplitude for the three point function with axion, electron and positron on the external legs is given by the sum of the tree level part, given in Eq.~\eqref{eq:Atree}, and the one-loop contribution given in Eq.~\eqref{eq:loopamp3}. The sum explicitly reads
\be
i \, \mathcal{A} = i \, \mathcal{A}_{\rm tree} + i \, \mathcal{A}_{\rm loop} =  \frac{1}{2 f} \, k_\mu \, \bar{u}(p_1) \gamma^\mu \gamma^5 v(p_2)  \left[ c^0_e - c^0_\ell \; \lambda_\ell^2 \; \frac{4 g_A^\ell g_A^e}{16 \pi^2 \epsilon} \right] \ .
\label{eq:fullamplitude}
\ee
We make this amplitude finite through renormalization of the couplings, employing a mass-independent subtraction scheme. The coefficients $c^0_e$ and $c^0_\ell$ appearing in Eq.~\eqref{eq:fullamplitude} are infinite bare couplings, and only the renormalization of $c^0_e$ is required to make the full amplitude finite
\be
c_e^0 = 
(1 + \delta Z_{e \ell}) c_e \ , \qquad \qquad \qquad \delta Z_{e \ell} = \frac{c_\ell}{c_e} \; \lambda_\ell^2 \; \frac{4 g_A^\ell g_A^e}{16 \pi^2 \epsilon} \ ,
\label{eq:Z}
\ee
The evolution for the renormalized coupling $c_e$ is obtained by imposing that the bare coupling $c_e^0$ does not depend on the renormalization scale $\mu$. Upon employing standard techniques~\cite{Buras:1998raa}, we find the renormalization group (RG) equation for the renormalized coupling
\be
\frac{d c_e}{d \ln \mu} = \frac{8 \lambda_\ell^2 g_A^\ell g_A^e}{16 \pi^2} \; c_\ell = \frac{\lambda_\ell^2}{8 \pi^2} \; c_\ell \ ,
\ee
where in the last step we have used the relations $g_A^\ell = g_A^e = +1/2$. The physics behind the coupling evolution described by the above differential equation is identical to the one associated to mixing among axial-vector currents of SM fermions coupled to WIMP DM, first pointed out in Refs.~\cite{Crivellin:2014qxa,DEramo:2014nmf}. A complete RG treatment of the problem is not necessary in our case, since the lepton Yukawa couplings are much smaller than one and we never need to resum large logarithms. Instead, we approximately solve the above RG equation, ignoring the energy dependence of the lepton Yukawa coupling, and we find 
\be
c_e = c_e(\Lambda) - \frac{\lambda_\ell^2}{8 \pi^2}  c_\ell \log\left(\Lambda / m_\ell \right) \ .
\ee
This solution, corresponding to the result of a fixed-order calculation, is in agreement with the analytical solutions provided in Refs.~\cite{DEramo:2016gos,DEramo:2017zqw}.

We conclude this Appendix with the derivation of the bounds in Eqs.~\eqref{eq:loopboundtau} and \eqref{eq:loopboundmu}. Assuming that the axion does not couple to electrons at high energy ({\rm i.e.}, $c_e(\Lambda) = 0$),  the bound in Eq.~\eqref{c_e} translates into the inequality 
\be
f / c_\ell \gtrsim  \;  \frac{\lambda_\ell^2}{8 \pi^2} \log\left(\Lambda / m_\ell \right) \; \times \; 5 \times 10^9 \, {\rm GeV} \ .
\ee
In order to translate the bound above into a constraint on $f/c_\ell$, we need to choose a value for the cutoff scale $\Lambda$. We assume an order one coupling $c_\ell \simeq \mathcal{O}(1)$ and we identify the cutoff scale with $f$, finding
\begin{eqnarray}
f/c_\tau & \gtrsim & 7 \times 10^4 \, {\rm GeV} \, , \\
f/c_\mu & \gtrsim & 200  \, {\rm GeV} \ .
\end{eqnarray}

\section{Cross sections, decay widths and thermal averages}
\label{app:calculations}

In this Appendix, we give details about the evaluation of cross sections and decay widths.~\footnote{See also Refs.~\cite{Turner:1986tb,Brust:2013xpv,Baumann:2016wac} for previous estimations and~\cite{Ferreira:2018vjj} for the calculation of the cross sections in the case of quarks.} After presenting our calculations, we also provide a general equation to compute thermally averaged cross sections.

\subsection*{Scattering cross sections}

For a scattering such as those in Fig.~\ref{fig:FeynmanScattering}, we introduce the Mandelstam variables
\begin{align}
s = & \, (p_1 + p_2)^2 = (p_3 + k)^2  \ , \\
t = & \, (p_1 - p_3)^2 = (p_2 - k)^2  \ , \\
u = & \, (p_1 - k)^2 = (p_2 - p_3)^2  \ ,
\end{align}
where we denote with $p_i$ and $k$ the four-momenta of the bath particle $B_i$ and of the axion, respectively. Conservation of four-momentum implies the constraint $s + t + u = \sum_i m_i^2$. The Lorentz invariant cross section is defined as
\be
\sigma_{B_1 B_2 \, \rightarrow \, B_3 a} = \frac{1}{4 I} \, \int \, \left|\mathcal{M}_{B_1 B_2 \, \rightarrow \, B_3 a}\right|^2 \, d \Phi^{(2)} \ .
\label{eq:CSgeneral}
\ee
The flux factor $I$ is defined in terms of the initial four-momenta and it can be written in a manifest Lorentz invariant form
\be
I = \sqrt{(p_1 \cdot p_2)^2 - m_1^2 m_2^2} = \frac{s}{2} \, \sqrt{1 - \frac{2 (m_1^2 + m_2^2)}{s} + \frac{(m_1^2 - m_2^2)^2}{s^2}}   \ .
\ee
The Lorentz invariant phase space reads
\be
d \Phi^{(2)} = (2 \pi)^4 \delta^4(p_1 + p_2 - p_3 - k) \frac{d^3 p_3}{2 E_3 (2 \pi)^3} \frac{d^3 k}{2 E_k (2 \pi)^3} \ .
\label{eq:dPhi2}
\ee
The only remaining quantities to compute before integrating over the phase space are the squared matrix elements. We take them averaged and summed over initial and final polarizations, respectively. Once we have completed this task, it is straightforward to determine the scattering cross section, and we will compute the explicit form for each process analyzed in this paper. Since the cross section ultimately depends only on $s$, we can perform the phase space integration in any frame, as long as we express the final result in a manifest Lorentz invariant form. The center of mass (CM) frame is particularly convenient for this operation, and in this frame the differential phase space results in
\be
\left. d \Phi^{(2)}\right|^{\rm CM} = \frac{d \Omega}{32 \pi^2} \left( 1 - \frac{m_3^2}{s} \right) \ ,
\label{eq:dPhi2CM}
\ee
where we neglect the axion mass in the final state.

\subsubsection*{Cross Section for $\ell^+ \ell^- \rightarrow \gamma a$}

\begin{figure}
\begin{center}		
\includegraphics[width=0.95\linewidth]{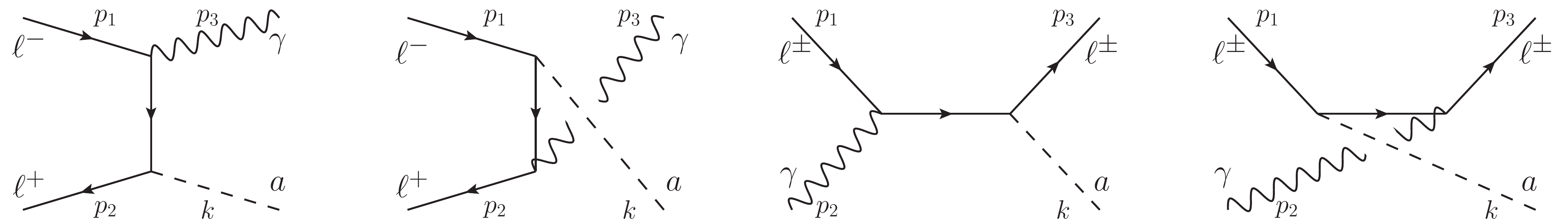}
\caption{Feynman diagrams for axion production via scattering. From left to right: lepton pair annihilation (t- and u-channel), Compton-like scattering (s- and u-channel).}
\label{fig:FeynmanScattering}
\end{center}
\end{figure}

The matrix element for this process is the sum of two contributions
\be
\mathcal{M}_{\ell^+ \ell^- \rightarrow \gamma a} = \mathcal{M}_{\ell^+ \ell^- \rightarrow \gamma a}^{(t)} + \mathcal{M}^{(u)}_{\ell^+ \ell^- \rightarrow \gamma a} \ ,
\ee
corresponding to t- and u-channel exchange (first two Feynman diagrams in Fig.~\ref{fig:FeynmanScattering}). Their explicit expressions read
\begin{align}
i \, \mathcal{M}_{\ell^+ \ell^- \rightarrow \gamma a}^{(t)}  = & \,  \frac{c_\ell \, e}{2 f} \; \epsilon^*_\mu(p_3, \lambda) \, k_\nu \; \bar{v}(p_2) \left( \gamma^\nu \gamma^5 \right) \, \frac{i (\slashed{p_1} - \slashed{p_3} + m_\ell)}{t - m_\ell^2} \left( i \, \gamma^\mu \right)  u(p_1)  \ ,  \\ 
i \, \mathcal{M}_{\ell^+ \ell^- \rightarrow \gamma a}^{(u)}  = &\,  \frac{c_\ell \, e}{2 f} \; \epsilon^*_\mu(p_3, \lambda) \, k_\nu \;  \bar{v}(p_2) \,\left( i \, \gamma^\mu \right)\, \frac{i (\slashed{p_1} - \slashed{k} + m_\ell)}{u - m_\ell^2}   \left( \gamma^\nu \gamma^5 \right)  u(p_1)  \ . 
\end{align}
Here, $e$ is the electric charge of the lepton (they are all equal to the one for the electron). The polarization tensor $\epsilon_\mu(p_3, \lambda)$ accounts for the production of a photon with helicity $\lambda$ and four-momentum $p_3$. We sum the two contributions, and compute the squared amplitude
\be
\left|\mathcal{M}_{l^+ l^- \rightarrow \gamma a} \right|^2 = \frac{1}{4} \, \times \, \frac{c^2_\ell \, e^2}{4 f^2} \,  (- g_{\mu\alpha}) \, k_\nu k_\beta \, \mathcal{D}_1^{\mu\nu\alpha\beta}  \ , 
\ee
where the tensor with the trace over Dirac indices in this case reads
\be
\begin{split}
\mathcal{D}_1^{\mu\nu\alpha\beta} =  {\rm Tr} & \, \left[ \left( \slashed{p_1} + m_\ell \right) \left(\frac{\gamma^\alpha \left(\slashed{p_1} - \slashed{p_3} + m_\ell \right) \gamma^\beta \gamma^5 }{t - m^2_\ell}   + \frac{\gamma^\beta \gamma^5 \left(\slashed{p_1} - \slashed{k} + m_\ell \right) \gamma^\alpha}{u - m^2_\ell}  \right) \right.\times \\ & \left.\left( \slashed{p_2} - m_\ell \right)  \left(\frac{\gamma^\nu \gamma^5 \left(\slashed{p_1} - \slashed{p_3} + m_\ell \right) \gamma^\mu}{t - m^2_\ell}   + \frac{\gamma^\mu \left(\slashed{p_1} - \slashed{k} + m_\ell \right) \gamma^\nu \gamma^5 }{u - m^2_\ell}   \right) \right] \ .
\end{split}
\ee
Notice how we average over the initial $2 \times 2 = 4$ polarizations. After straightforward Dirac algebra, and after putting the external particles on-shell, the squared matrix element can be expressed in terms of the Mandelstam variables as
\be
\left|\mathcal{M}_{\ell^+ \ell^- \rightarrow \gamma a} \right|^2 = \frac{c_\ell^2 e^2}{f^2} \frac{m_\ell^2 \, s^2}{(m_\ell^2 - t) (s + t - m_\ell^2)} \ .
\label{eq:squaredMpair}
\ee

It is convenient to perform the phase-space integration in the CM frame and express the final result in a manifest Lorentz invariant form. Let  $\theta$ be the scattering angle in such a frame. The relation between the Mandelstam variable $t$ and $\theta$ is given by
\be
\left. t \right|_{\ell^+ \ell^- \rightarrow \gamma a}= m_\ell^2 - \frac{s}{2} \left[1 - \sqrt{1 - \frac{4 m_\ell^2}{s}} \cos\theta \right]  \ .
\ee
We plug this expression for $t$ into the squared matrix element in Eq.~\eqref{eq:squaredMpair}. The total cross section can be obtained from the general expression in Eq.~\eqref{eq:CSgeneral}. After integrating over the phase space, which for this process reads
\be
\left. d \Phi^{(2)} \right|^{\rm CM}_{\ell^+ \ell^- \rightarrow \gamma a} = \frac{d \cos\theta}{16 \pi}  \ ,
\label{eq:dPhi2CMpair}
\ee
and accounting for the flux factor for this process
\be
\left. I \right|_{\ell^+ \ell^- \rightarrow \gamma a}  = \frac{\sqrt{s}}{2} \sqrt{1 - \frac{4 m_\ell^2}{s}}   \ ,
\ee
we find the total cross section
\begin{equation}
\sigma_{\ell^+ \ell^- \rightarrow \gamma a}(s) = \frac{c_\ell^2 e^2}{f^2} \frac{m_\ell^2 \tanh^{-1}\left( \sqrt{1 - \frac{4 m_\ell^2}{s}}\right)}{4 \pi   (s - 4 m_\ell^2)} \ ,
\label{eq:sigmapair}
\end{equation}
written in a manifest Lorentz invariant form. 

\subsubsection*{Cross Section for $\ell^\pm \gamma \rightarrow \ell^\pm a$}

This process also gets two contributions
\be
\mathcal{M}_{\ell^\pm \gamma \rightarrow \ell^\pm a} = \mathcal{M}_{\ell^\pm \gamma \rightarrow \ell^\pm a}^{(s)} + \mathcal{M}^{(u)}_{\ell^\pm \gamma \rightarrow \ell^\pm a} \ ,
\ee
in this case associated to s- and u-channel exchange (third and fourth Feynman diagrams in Fig.~\ref{fig:FeynmanScattering}). Their explicit expressions read
\begin{align}
i \, \mathcal{M}_{\ell^\pm \gamma \rightarrow \ell^\pm a}^{(s)} = & \,  \frac{c_\ell \, e}{2 f} \; \epsilon_\mu(p_2, \lambda) \, k_\nu \; \bar{u}(p_3) \left( \gamma^\nu \gamma^5 \right) \, \frac{i (\slashed{p_1} - \slashed{p_2} + m_\ell)}{s - m_\ell^2} \left( i \, \gamma^\mu \right)  u(p_1)   \, ,\\ 
i \, \mathcal{M}^{(u)}_{\ell^\pm \gamma \rightarrow \ell^\pm a}   = & \,  \frac{c_\ell \, e}{2 f} \; \epsilon_\mu(p_2, \lambda)  \, k_\nu \; \bar{u}(p_3) \,\left( i \, \gamma^\mu \right)\, \frac{i (\slashed{p_1} - \slashed{k} + m_\ell)}{u - m_\ell^2}   \left( \gamma^\nu \gamma^5 \right)  u(p_1)  \, . \\ 
\end{align}
The squared amplitude can be again written as
\be
\left|\mathcal{M}_{\ell^\pm \gamma \rightarrow \ell^\pm a} \right|^2 = \frac{1}{4} \, \times \, \frac{c^2_\ell \, e^2}{4 f^2} \,  (- g_{\mu\alpha}) \, k_\nu k_\beta \, \mathcal{D}_2^{\mu\nu\alpha\beta}  \ , 
\ee
where this time the Dirac part takes the form
\be
\begin{split}
\mathcal{D}_2^{\mu\nu\alpha\beta} =  {\rm Tr} & \, \left[ \left( \slashed{p_1} + m_\ell \right) \left(\frac{\gamma^\alpha \left(\slashed{p_1} + \slashed{p_2} + m_\ell \right) \gamma^\beta \gamma^5 }{s - m^2_\ell}   + \frac{\gamma^\beta \gamma^5 \left(\slashed{p_1} - \slashed{k} + m_\ell \right) \gamma^\alpha}{u - m^2_\ell}  \right) \right.\times \\ & \left.\left( \slashed{p_3} + m_\ell \right)  \left(\frac{\gamma^\nu \gamma^5 \left(\slashed{p_1} + \slashed{p_2} + m_\ell \right) \gamma^\mu}{s - m^2_\ell}   + \frac{\gamma^\mu \left(\slashed{p_1} - \slashed{k} + m_\ell \right) \gamma^\nu \gamma^5 }{u - m^2_\ell}   \right) \right] \ .
\end{split}
\ee
After the usual Dirac algebra and putting external legs on-shell, we find
\be
\left|\mathcal{M}_{\ell^\pm \gamma \rightarrow \ell^\pm a} \right|^2 = \frac{c_\ell^2 e^2}{f^2} 
\frac{m_\ell^2 \, t^2}{(s- m_\ell^2) (s + t - m_\ell^2)} \ .
\label{eq:squaredMCompton}
\ee
This result satisfies a crossing symmetry property, which we can use as a check of our calculation. If we look at Feynman diagrams, we see that to go from pair annihilation to scattering we need the following replacements in the matrix elements
\be
p_1 \rightarrow p_1 \ , \qquad p_2 \rightarrow - p_3 \ , \qquad p_3 \rightarrow - p_2 \ , \qquad k \rightarrow k \ .
\ee
This in turn implies the replacements for the Mandelstam variables
\be
s \; \rightarrow \; t \, \qquad \qquad t \; \rightarrow \; s \, \qquad \qquad u \; \rightarrow \; u \ .
\ee
If we plug this crossing symmetry transformation in Eq.~\eqref{eq:squaredMpair}, and we account for an overall minus sign since we are crossing one fermion field, we find the expression in Eq.~\eqref{eq:squaredMCompton}.

The phase space integration can be again performed in the CM frame, where the different kinematics leads to the following relation between $t$ and the scattering angle
\be
\left. t \right|_{\ell^\pm \gamma \rightarrow \ell^\pm a} = - \frac{s}{2} \left( 1 - \frac{m_\ell^2}{s} \right)^2 \left( 1 - \cos\theta \right)  \ .
\ee
We integrate the resulting squared matrix element over the phase space for this process
\be
\left. d \Phi^{(2)} \right|^{\rm CM}_{\ell^\pm \gamma \rightarrow \ell^\pm a}  = \frac{d \cos\theta}{16 \pi} \left( 1 - \frac{m_\ell^2}{s} \right)  \ .
\label{eq:dPhi2CMcompton}
\ee
After we account for the flux factor,
\be
\left. I \right|^{\rm CM}_{\ell^\pm \gamma \rightarrow \ell^\pm a}   = \frac{\sqrt{s}}{2} \left( 1 - \frac{m_\ell^2}{s} \right) \ ,
\ee
and we find the Lorentz invariant cross section for this process
\begin{equation}
\sigma_{\ell^\pm \gamma \rightarrow \ell^\pm a}(s) = \frac{c_\ell^2 e^2}{f^2} \frac{m_\ell^2 [2 s^2 \log(s/m_\ell^2) - 3 s^2 + 4 m_\ell^2 s - m_\ell^4]}{32 \pi s^2 (s - m_\ell^2)} \ .
\label{eq:sigmacompton}
\end{equation}

\subsection*{Decay widths}

The decay width calculation goes along similar lines. The decay rate is defined as 
\be
\Gamma_{B_1 \rightarrow \, B_3 a} = \frac{1}{2 m_1} \,\int \left|\mathcal{M}_{B_1 \, \rightarrow \, B_3 a}\right|^2 \, d \Phi^{(2)} \ ,
\ee
where the two-body phase space is still given by the expression in Eq.~\eqref{eq:dPhi2}, which becomes Eq.~\eqref{eq:dPhi2CM} in the rest frame of the decaying particle. All we need is again the squared matrix element, and then we integrate over the possible final states.

We compute the decay width for the decay processes $\ell^\pm \, \rightarrow \, \ell^{\prime \pm}  a$ mediated by the off-diagonal lepton currents. The matrix element reads
\be
i \, \mathcal{M}_{l^- \rightarrow l^{- \prime} a} = \frac{k_\mu}{2 f} \, \bar{u}(p) \gamma^\mu \left(\mathcal{V}_{\ell^\prime \ell} + \mathcal{A}_{\ell^\prime \ell} \gamma^5 \right) u(q) \ .
\ee
where we impose four-momentum conservation in the form $q = p + k$, and $k$ is the four-momentum of the axion. We square this matrix element, and we average over the two initial polarizations of the decaying particle
\be
\left| \mathcal{M}_{\ell^\pm \, \rightarrow \, \ell^{\prime \pm}  a} \right|^2 = \frac{k_\mu k_\nu}{8 f^2}  {\rm Tr}
\left[ \left( \slashed{q} + m_{l} \right)  \gamma^\nu  \left(\mathcal{V}_{\ell^\prime \ell} + \mathcal{A}_{\ell^\prime \ell} \gamma^5 \right)  \left( \slashed{p} + m_{l^\prime} \right) \gamma^\mu \left(\mathcal{V}_{\ell^\prime \ell} + \mathcal{A}_{\ell^\prime \ell} \gamma^5 \right) \right] \ .
\ee
After performing the Dirac algebra, Lorentz contractions and putting the external states on-shell, we find
\be
\left| \mathcal{M}_{\ell^\pm \, \rightarrow \, \ell^{\prime \pm}  a} \right|^2 = \left( \mathcal{V}^2_{\ell^\prime \ell} + \mathcal{A}^2_{\ell^\prime \ell}  \right) \frac{m_\ell^4}{4 f^2}  \left( 1 - \frac{m_{\ell^\prime}^2}{m_\ell^2} \right)^2 \ .
\ee
The phase space integral is straightforward, and it ends up giving a factor
\be
\Phi^{(2)} = \frac{1}{8 \pi} \left( 1 - \frac{m_{\ell^\prime}^2}{m_\ell^2} \right) \ . 
\ee
Thus the decay width results in
\be
\Gamma_{l^- \rightarrow l^{- \prime} a} = \left( \mathcal{V}^2_{\ell^\prime \ell} + \mathcal{A}^2_{\ell^\prime \ell}  \right)  \frac{m_\ell^3}{64 \pi f^2} \left( 1 - \frac{m_{\ell^\prime}^2}{m_\ell^2} \right)^3 \ .
\ee

\begin{figure}
	\centering
	\includegraphics[width=0.43\linewidth]{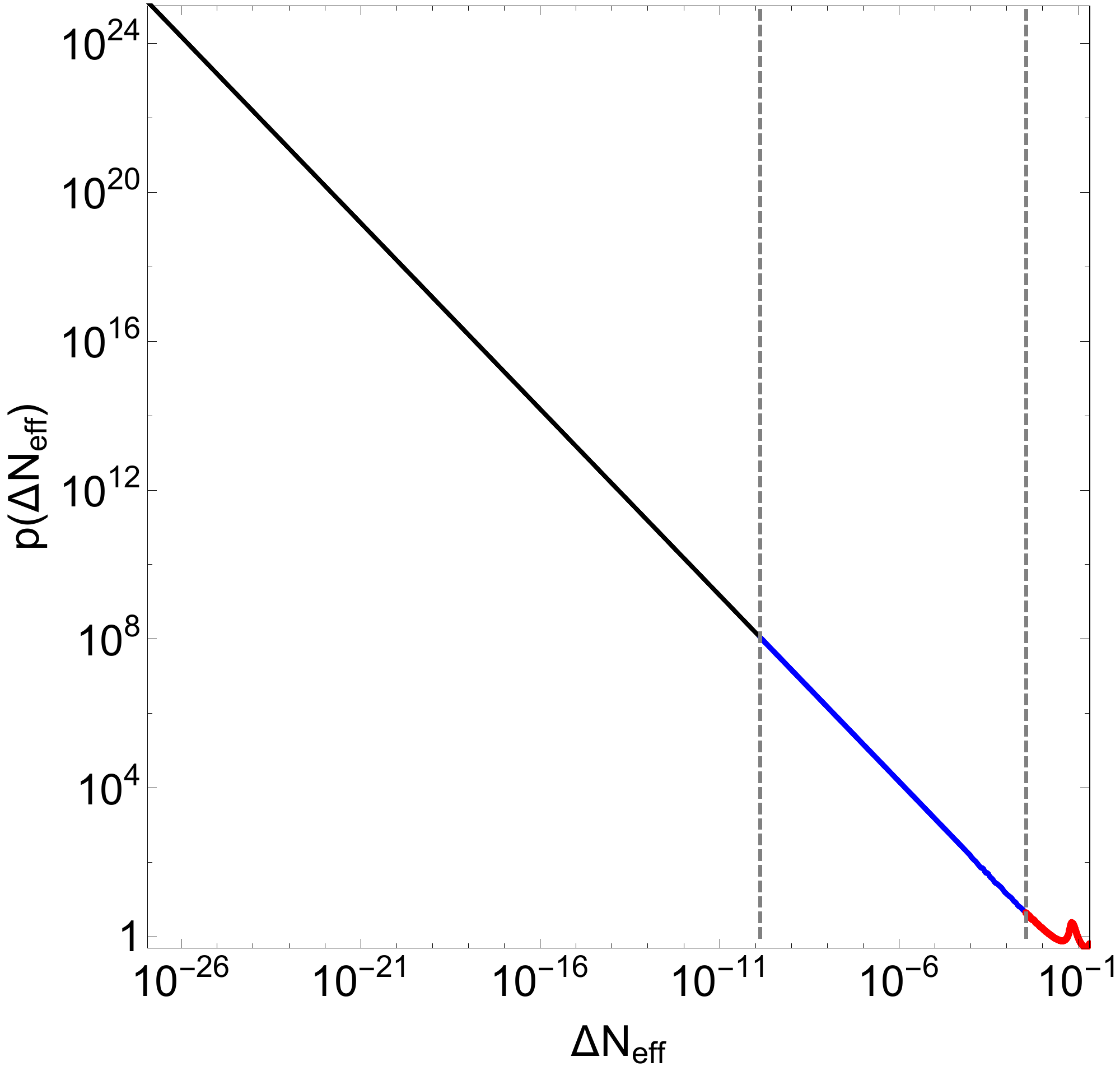} $\qquad$
	\includegraphics[width=0.41\linewidth]{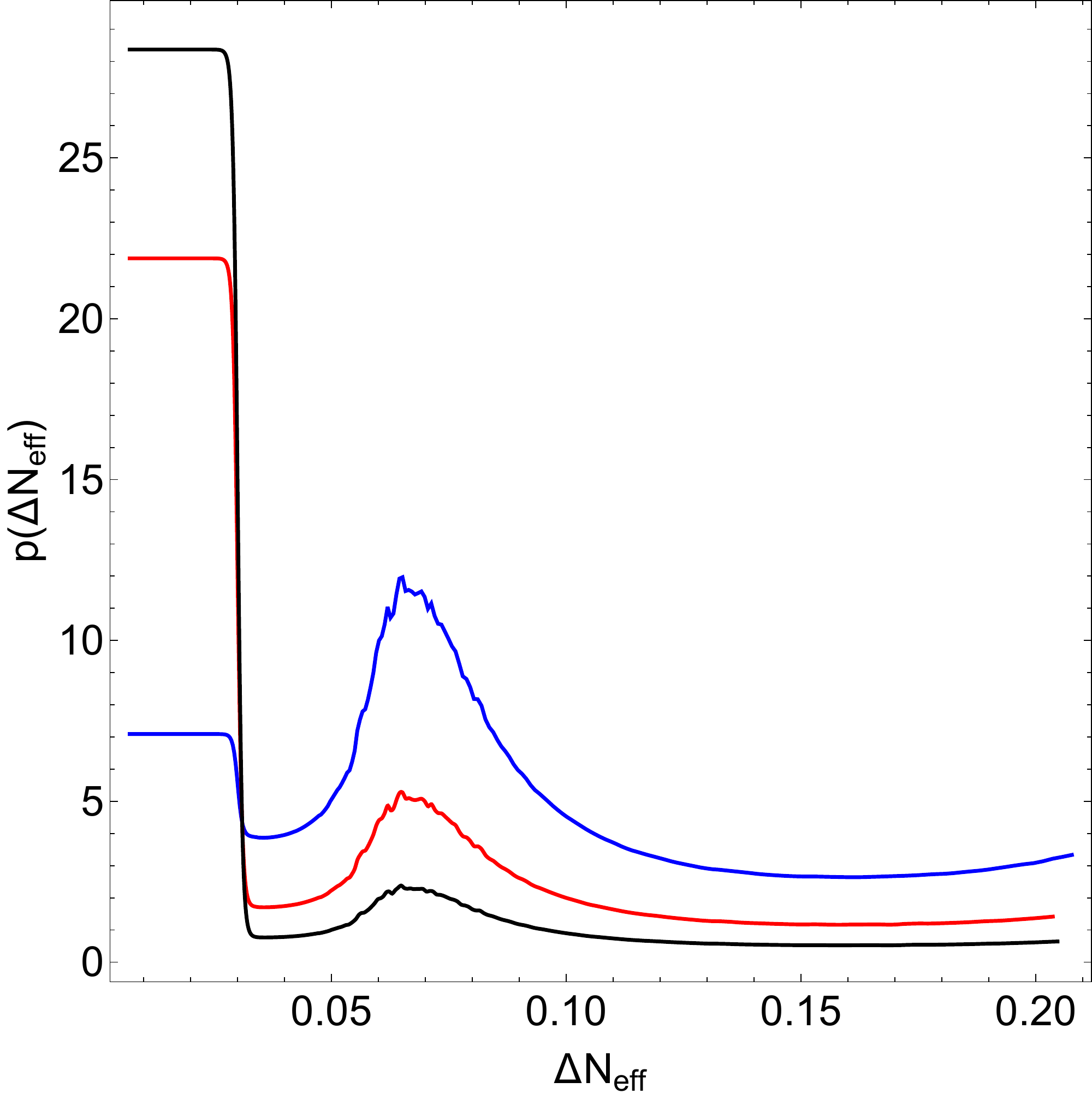}
	\includegraphics[width=0.43\linewidth]{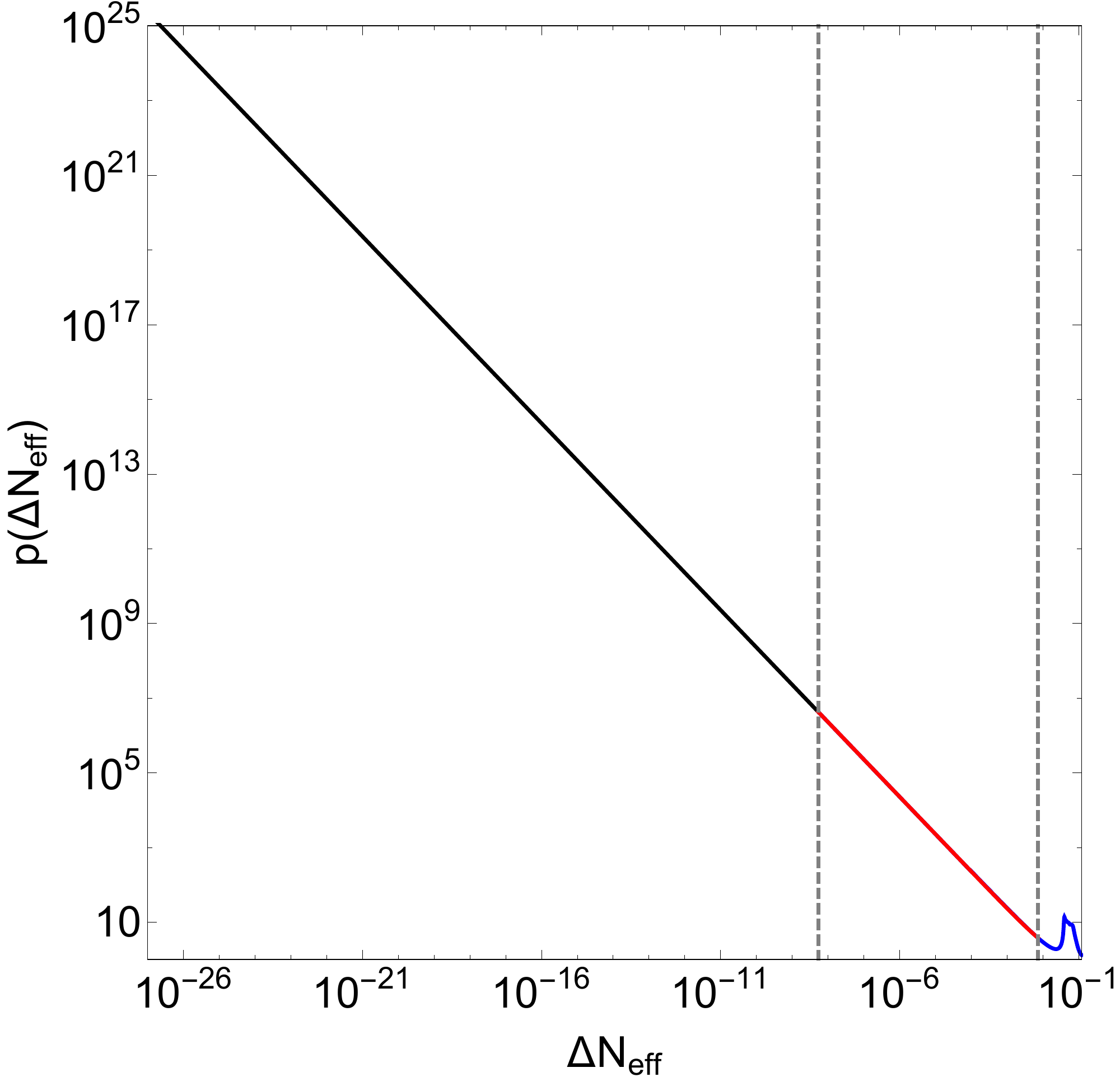} $\qquad$
	\includegraphics[width=0.41\linewidth]{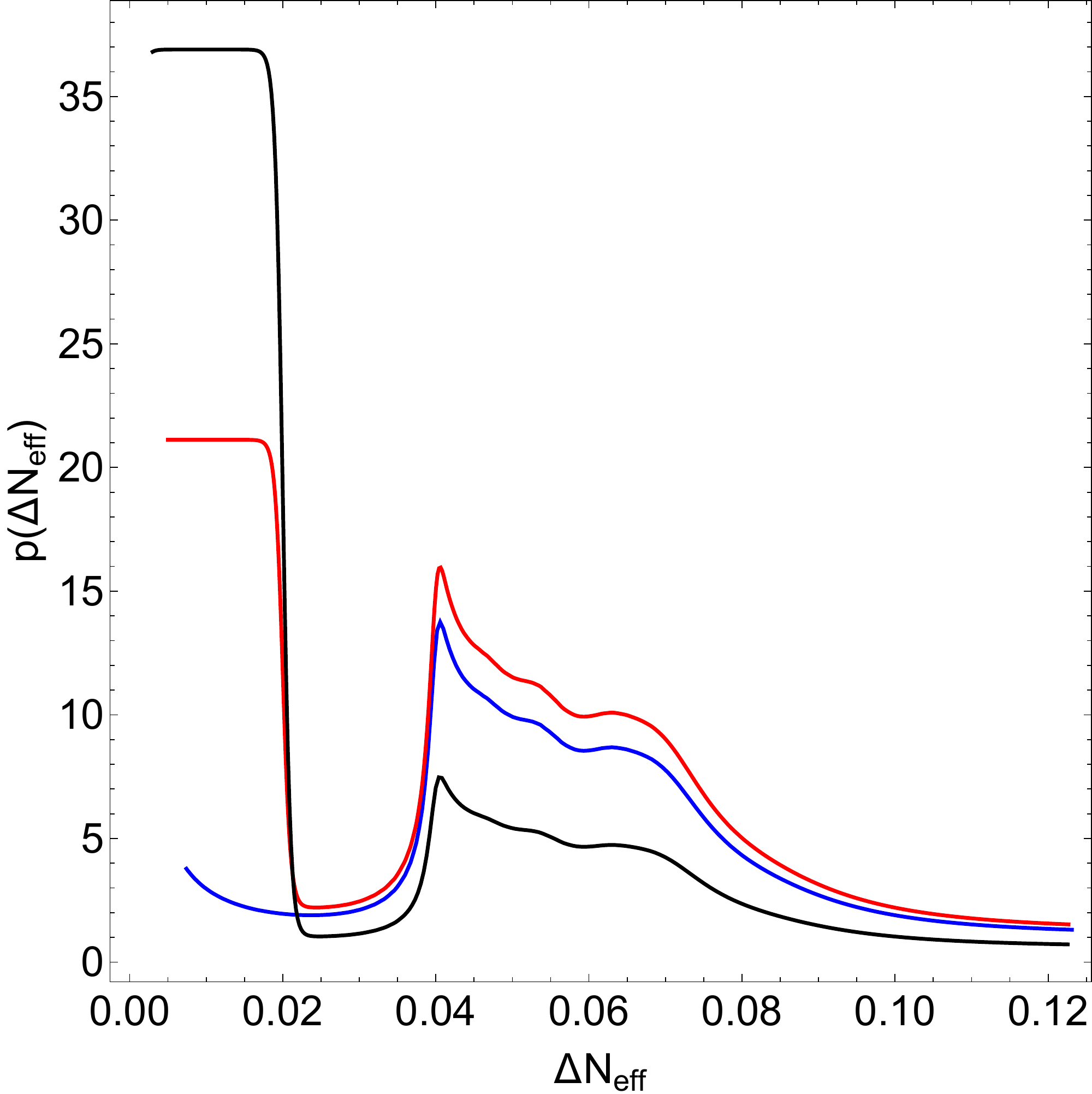}
	\includegraphics[width=0.43\linewidth]{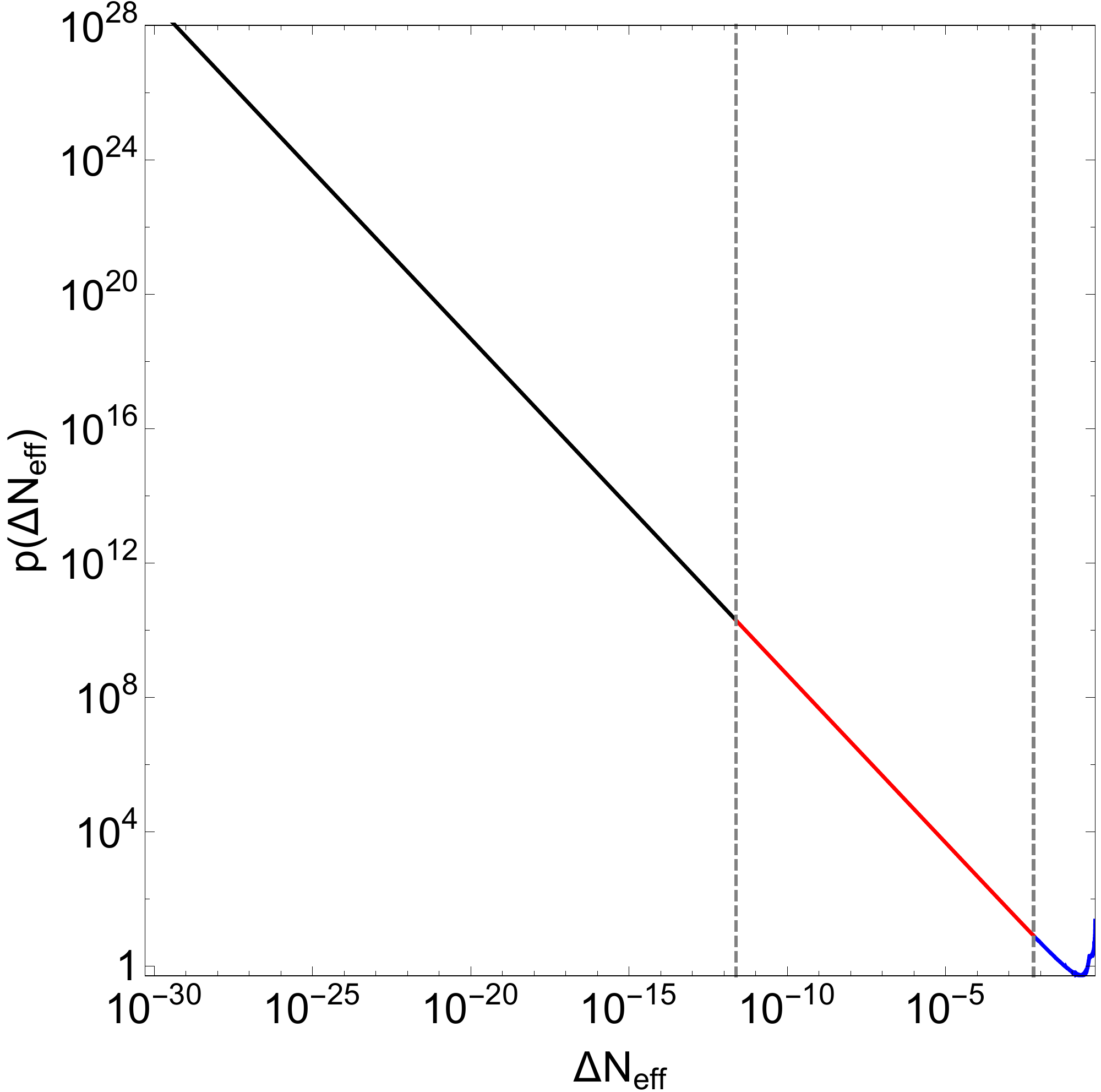} $\qquad$
	\includegraphics[width=0.41\linewidth]{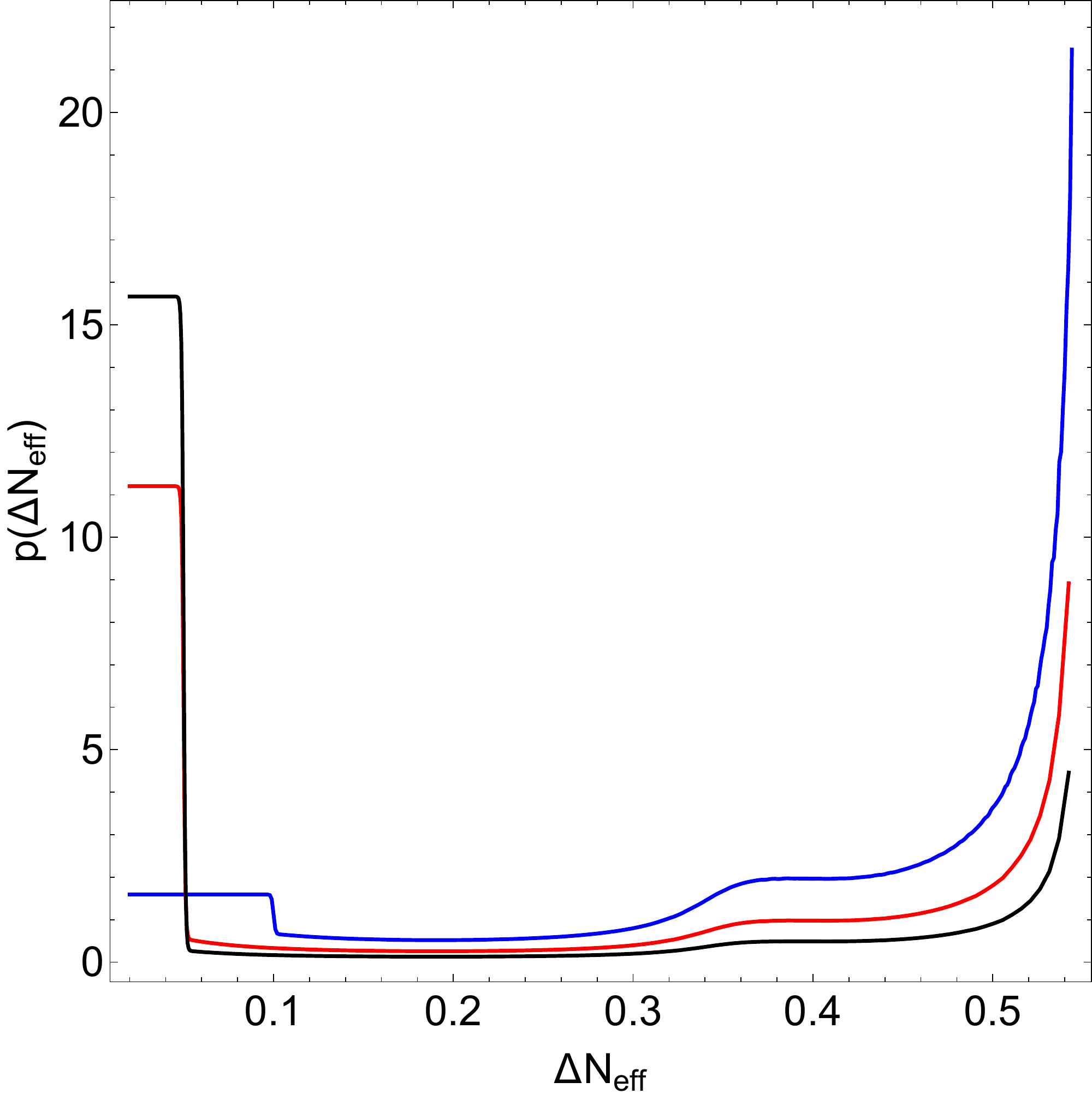}
	\caption{ Prior probability density as a function of $\dN$. The first line corresponds to $\tau$-decay, the second to $\tau$-scatterings and the third to $\mu$-scatterings. The plots on the left are the original priors and the ones on the right are the modified priors that we used in the runs. The colors correspond to different values of $(f/c_\ell)|_{max}$ considered: (black, $10^{18}$GeV), (red, $10^{11}$GeV) and (blue, $\sim10^{8}$GeV).
		\label{fig:priors}}
\end{figure}

\subsection*{Thermally averaged cross sections}

The scattering rate appearing in the Boltzmann equation, as defined in Eq.~\eqref{eq:collscattering}, has a thermal average of the total cross section times the Moeller velocity. Here, we provide a general equation that allows us to compute this quantity valid within the Boltzmann approximation for the phase space distribution functions, $f(E) = \exp[ - E / T]$. Once we compute the Lorentz invariant expression for the total cross section, the thermal average reads~\cite{Giudice:2003jh,DEramo:2017ecx}
\be
\langle \sigma_{B_1 B_2 \rightarrow B_3 a} v \rangle =  \frac{\int_{s^{\rm min}_{12}}^\infty ds \; \lambda(s, m_{B_1}, m_{B_2}) \; s^{-1/2} \; \sigma_{B_1 B_2 \rightarrow B_3 \chi}(s)  K_1[\sqrt{s} / T]}{8 \, K_2[m_{B_1} / T] K_2[m_{B_2} / T] \, m_{B_1}^2 m_{B_2}^2 T} \ .
\label{eq:thermaverag}
\ee
Here, the function $\lambda$ and the  lower integration extremum are defined as follows
\begin{align}
\label{eq:lambdadef}  \lambda(x, y, z) \equiv & \, [x - (y + z)^2] [x - (y - z)^2]  \ , \\
s^{\rm min}_{12} \equiv & \, (m_{B_1} + m_{B_2})^2 \ .
\end{align}
The expression in Eq.~\eqref{eq:thermaverag} is valid when both initial state particles are massive. In this work, we have one process where the photon is in the initial state. For this case, when $m_{B_2} = 0$, the thermal average results in
\be
\left. \langle \sigma_{B_1 B_2 \rightarrow B_3 a} v \rangle \right|_{m_{B_2} = 0} =  \frac{\int_{m_{B_1}^2}^\infty ds \; \lambda(s, m_{B_1}, 0) \; s^{-1/2} \; \sigma_{B_1 B_2 \rightarrow B_3 \chi}(s)  K_1[\sqrt{s} / T]}{16 \, K_2[m_{B_1} / T] \, m_{B_1}^2 T^3}  \ .
\label{eq:thermaverag2}
\ee

\section{Priors on $\dN$ from flat prior on $\log_{10}(f/c_\ell)$}\label{App:Cosmo}

In this final Appendix, we give explicit results for the priors on $\dN$ as they result from our choice of flat priors on $\log_{10}(f/c_\ell)$. The relation between $\log_{10}(f/c_\ell)$ and $\dN$ for the different cases studied in this work can be found in Fig.~\ref{figura}. As discussed in Sec.~\ref{sec:cosmo}, we consider three different upper bounds for $(f/c_{\ell})|_{\rm max} = \left(10^{18}, 10^{11}, 10^8\right)$ GeV. The resulting prior probability densities $p(\Delta N_{\rm eff})$ are shown on the left column of Fig.~\ref{fig:priors} for each case and for different upper bounds on $f/c_\ell$. Note that large tails at $\dN\sim 0$ imply a significant fraction of the whole cumulative probability of the prior in that region. Indeed at large $f/c_\ell$ we know that $\dN\propto (f/c_\ell)^{-8/3}$ and this implies that $p(\Delta N_{\rm eff})\propto 1/\Delta N_{\rm eff}$.  The interval $\dN\sim 0$ in the parameter space is however very difficult to sample practically and so, taking into account that the data is almost insensitive to variations in $\dN$ below 0.01, we slightly modify our prior: we use a step-like feature at very small $\dN$ in order to have a compromise between the cumulative probability of the prior and the technical issue of the sampling at low $\dN$. The modified priors that we use are shown in the right column of Fig.~\ref{fig:priors}.

\bibliographystyle{JHEP}
\bibliography{Accione}

\end{document}